\documentclass[lettersize,journal]{IEEEtran}
\usepackage{amsmath,amssymb,amsfonts}
\usepackage{array}
\usepackage{bbding}

\usepackage{multirow}
\usepackage[caption=false,font=normalsize,labelfont=sf,textfont=sf]{subfig}
\usepackage{bm}
\usepackage[ruled,vlined,linesnumbered]{algorithm2e}
\usepackage{textcomp}
\usepackage{stfloats}
\usepackage{url}
\usepackage{verbatim}
\usepackage{graphicx}
\usepackage{xcolor}
\usepackage{cite}
\usepackage{threeparttable}

\begin{document}

\title{Humas: A Heterogeneity- and Upgrade-aware Microservice Auto-scaling Framework in Large-scale Data Centers}

\author{Qin Hua,~Dingyu Yang,~Shiyou Qian,~\IEEEmembership{Member,~IEEE},~Jian Cao,~\IEEEmembership{Senior Member,~IEEE},~Guangtao Xue,~\IEEEmembership{Member,~IEEE}, and~Minglu Li,~\IEEEmembership{Fellow,~IEEE}
\thanks{{\it Corresponding author: Shiyou Qian.}}

\thanks{Qin Hua, Shiyou Qian, Jian Cao, and Guangtao Xue are with the Department of Computer Science and Engineering, Shanghai Jiao Tong University,
Shanghai 200240, China (e-mail: huaqin@sjtu.edu.cn, qshiyou@sjtu.edu.cn, cao-jian@sjtu.edu.cn, gt\_xue@sjtu.edu.cn). 

Dingyu Yang is with Alibaba Group, Hangzhou,
Zhejiang 311100, China (e-mail: dingyu.ydy@alibaba-inc.com).

Minglu Li is with the School
of Computer Science and Technology, Zhejiang Normal University, Jinhua,
Zhejiang 321004, China (e-mail: mlli@zjnu.edu.cn).
}
}

\markboth{Journal of \LaTeX\ Class Files,~Vol.~14, No.~8, August~2021}%
{Hua \MakeLowercase{\textit{et al.}}: Humas: A Heterogeneity- and Upgrade-aware Microservice Auto-scaling Framework in Large-scale Data Centers}

\IEEEpubid{0000--0000/00\$00.00~\copyright~2021 IEEE}

\maketitle

\begin{abstract}
An effective auto-scaling framework is essential for microservices to ensure performance stability and resource efficiency under dynamic workloads. As revealed by many prior studies, the key to efficient auto-scaling lies in accurately learning performance patterns, i.e., the relationship between performance metrics and workloads in data-driven schemes. However, we notice that there are two significant challenges in characterizing performance patterns for large-scale microservices. Firstly, diverse microservices demonstrate varying sensitivities to heterogeneous machines, causing difficulty in quantifying the performance difference in a fixed manner. Secondly, frequent version upgrades of microservices result in uncertain changes in performance patterns, known as pattern drifts, leading to imprecise resource capacity estimation issues. To address these challenges, we propose Humas, a heterogeneity- and upgrade-aware auto-scaling framework for large-scale microservices. Firstly, Humas quantifies the difference in resource efficiency among heterogeneous machines for various microservices online and normalizes their resources in standard units. Additionally, Humas develops a least-squares density-difference (LSDD) based algorithm to identify pattern drifts caused by upgrades. Lastly, Humas generates capacity adjustment plans for microservices based on the latest performance patterns and predicted workloads. The experiment results conducted on 50 real microservices with over 11,000 containers demonstrate that Humas improves resource efficiency and performance stability by approximately 30.4\% and 48.0\%, respectively, compared to state-of-the-art approaches. 
\end{abstract}


\section{Introduction}
\label{sec:intro}
\IEEEPARstart{C}urrently, a vast number of microservices \cite{sinan} have been extensively deployed within the data centers of industry leaders, such as Google \cite{google_trace}, Amazon \cite{amazon}, and Alibaba \cite{luo2021characterizing}. Typically, each microservice encompasses multiple instances in the form of containers to manage substantial workloads in parallel. This approach facilitates the auto-scaling of resource capacity for microservices through the initiation or termination of containers. Given the dynamic nature of workloads, the implementation of an auto-scaling framework that effectively mitigates resource over- and under-provision issues is crucial for microservices to uphold resource efficiency \cite{deep-scaling,Meta} and maintain a high quality of service (QoS) \cite{sinan}. 


Prior research on microservice auto-scaling can be broadly divided into two categories: reactive approaches \cite{autoscaleopt,FIRM} and proactive approaches \cite{Meta,deep-scaling,ASARSA}. 
Reactive approaches monitor service performance and adjust resource capacity when performance violations are detected. For example, AutoScaleOpt \cite{autoscaleopt} modifies microservice capacity after CPU utilization violates the upper or lower thresholds. However, reactive approaches are unable to prevent anomalies beforehand, thereby compromising the stability of service performance. In contrast, proactive approaches construct models to understand the performance patterns of microservices which characterize the relationship between workload and performance. These approaches periodically adjust resource capacity based on the identified patterns and the predicted workload. For example, 
several studies \cite{deep-scaling, Meta, sriraman2019softsku, rzadca2020autopilot} train deep-learning (DL) models to comprehend the CPU usage patterns of microservices, aiming to stabilize their CPU utilization at target levels.

\IEEEpubidadjcol

The factors influencing the performance patterns of microservices can be broadly divided into two groups: internal and external. Firstly, the primary internal factor is business logic. When microservices alter their behavior through version upgrades, their performance patterns may also undergo changes. However, 
efficiently adapting to the time-varying performance patterns for auto-scaling decisions remains an open challenge, as it is challenging to automatically quantify the impact of pattern changes in a timely and precise manner. This limitation may lead to a decline in QoS and even impact service availability. Secondly, external factors can be further subdivided into two sub-categories: performance interference and hardware heterogeneity. Many studies focus on mitigating the interference of co-located microservices \cite{PARTIES, Quasar}, which is beyond the scope of this work. Hardware heterogeneity may significantly impact the operational efficiency of microservices and should be explicitly considered to accurately adjust capacity.

Through an analysis of a three-month trace of microservices, a significant observation of version upgrades has been identified. 
There exists a notable level of uncertainty regarding the impact of version upgrades on performance patterns. Certain upgrades result in a substantial increase or decrease in the CPU usage of microservices, sometimes by more than $100\%$, leading to shifts in the CPU usage pattern. It is crucial to emphasize that the mere adjustment of capacity through the routine re-learning of performance patterns at fixed intervals \cite{rzadca2020autopilot,Meta} may prove inadequate in capturing changes in a timely manner. Although some studies \cite{sinan,zhou2022cushion} advocate triggering adaptive pattern learning through upgrade notifications from developers, the implementation of such solutions imposes a considerable burden of maintenance on auto-scaling frameworks within large-scale data centers. Therefore, the accurate and automated detection of pattern drifts across thousands of microservices in production is imperative for the realization of precise capacity adjustment.

Moreover, the quantification of performance disparities among microservices that process dynamic workload on heterogeneous hardware remains an unresolved issue. Given the widespread existence of hardware heterogeneity, e.g., diverse CPU models \cite{sriraman2019softsku,Meta,zhou2022cushion} and machine capacities \cite{HARMONY}, deploying microservices on different machines can lead to significant variations in terms of resource efficiency. Existing studies, e.g. Google compute units (GCU) \cite{borg,rzadca2020autopilot}, measure difference based on benchmark performance using micro-architecture metrics, e.g. cycles per instruction (CPI) \cite{zhang2013cpi2}. However, these approaches have two limitations: 1) different microservices exhibit varying sensitivities to hardware heterogeneity, making comprehensive characterization using benchmarks challenging; and 2) micro-architecture metrics are workload-intensity dependent and cannot adapt to dynamic workloads \cite{yi2020cpi}.

To address these problems, we propose Humas, a {\bf h}etergeneity- and {\bf u}pgrade-aware {\bf m}icroservice {\bf a}uto-{\bf s}caling framework for large-scale data centers. The framework aims to maintain the CPU utilization of microservices at target levels. Firstly, we introduce a heterogeneity normalizer that dynamically measures the resource efficiency difference among heterogeneous hardware for various microservices with dynamic workloads. Secondly, we propose an online mechanism that identifies changes in CPU usage patterns, referred to as pattern drifts. Thirdly, we develop a capacity adjuster that periodically devises resource adjustment plans for microservices based on predicted workload and the latest detected performance patterns. As far as we know, Humas is the first microservice capacity planning framework that takes these factors into consideration.

To evaluate the efficacy of Humas, we conduct simulations using a two-month trace of 50 production microservices and compare it with four auto-scaling approaches. The experiment results demonstrate that Humas enhances the resource efficiency and stability of CPU utilization by approximately 30.4\% and 48.0\%, respectively, while also reducing the performance violation rate by about 52.1\%. 

{The contributions of this paper are summarized as follows}:
\begin{itemize}

\item {We are the first to conduct an analysis of the upgrade behaviors exhibited by diverse large-scale microservices, offering a thorough understanding of the implications of service upgrades on capacity estimation.} 

 \item We introduce an auto-scaling framework called Humas specifically designed for microservices in production data centers, which takes into account the impacts of version upgrades and hardware heterogeneity on service performance patterns.

 \item {We propose a systematic drift detection mechanism to automatically identify performance pattern drifts and accurately capture the latest CPU usage patterns for precise capacity estimation.}
 
 \item {We devise a comprehensive model to quantify the resource efficiency difference across heterogeneous machines under dynamic workloads for diverse microservices.} 
 
 \item We implement a prototype\footnote{The code of Humas and example data can be accessed at: https://github.com/Humas2023/Humas} of Humas and evaluate its effectiveness through data-driven simulations involving $50$ large-scale microservices with over $11,000$ containers.
\end{itemize}

\begin {figure}[tb]
    \centering
    \begin{minipage}[t]{0.51\linewidth}
    \centering
    \includegraphics[scale=0.162]{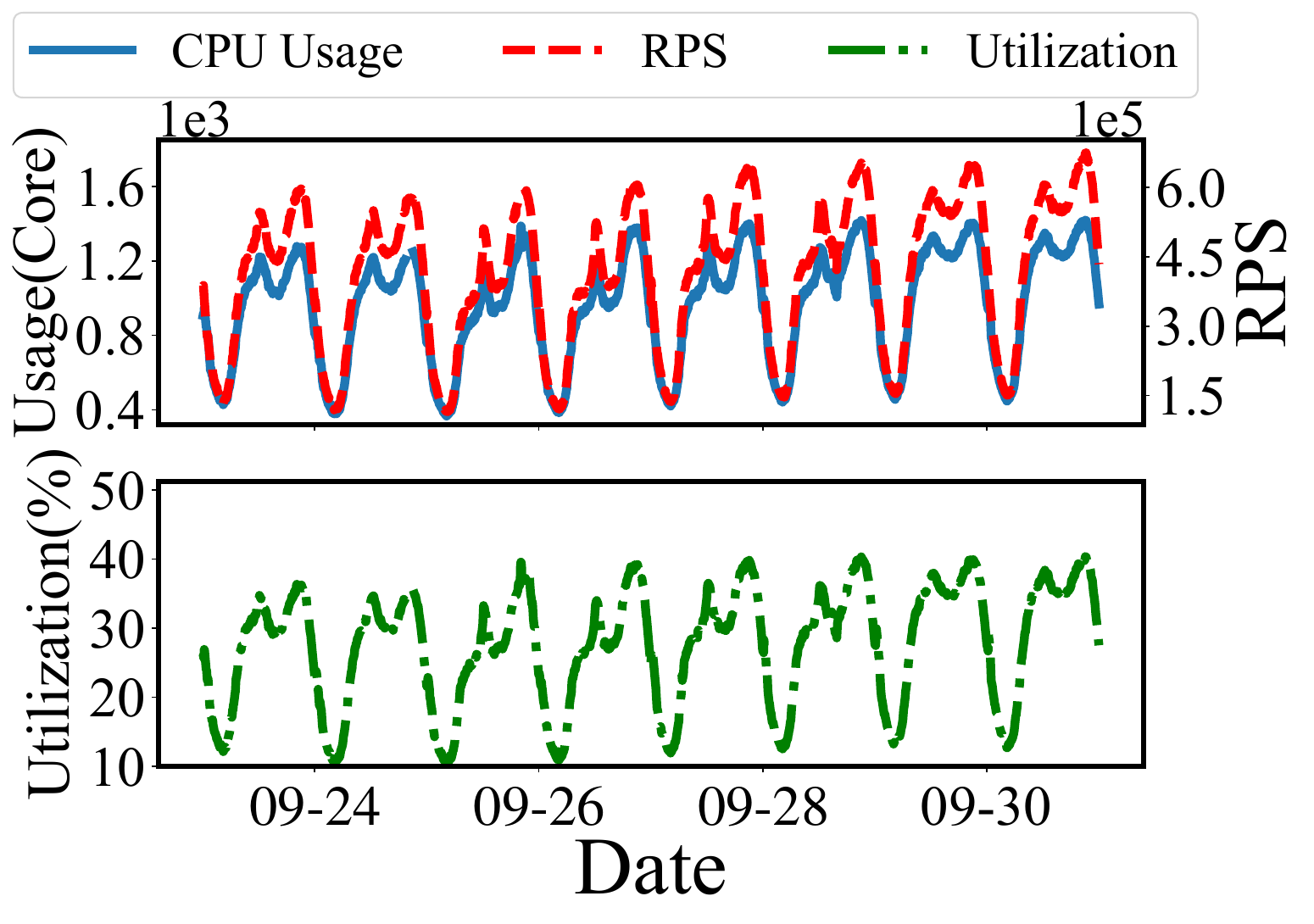}
    \caption{CPU usage and utilization of $MS_0$ with dynamic workload}
  \label{fig:util_usage_rps}
    \end{minipage}
    \hspace{0.1cm}
     \begin{minipage}[t]{0.445\linewidth}
  \centering
    \includegraphics[scale=0.16]{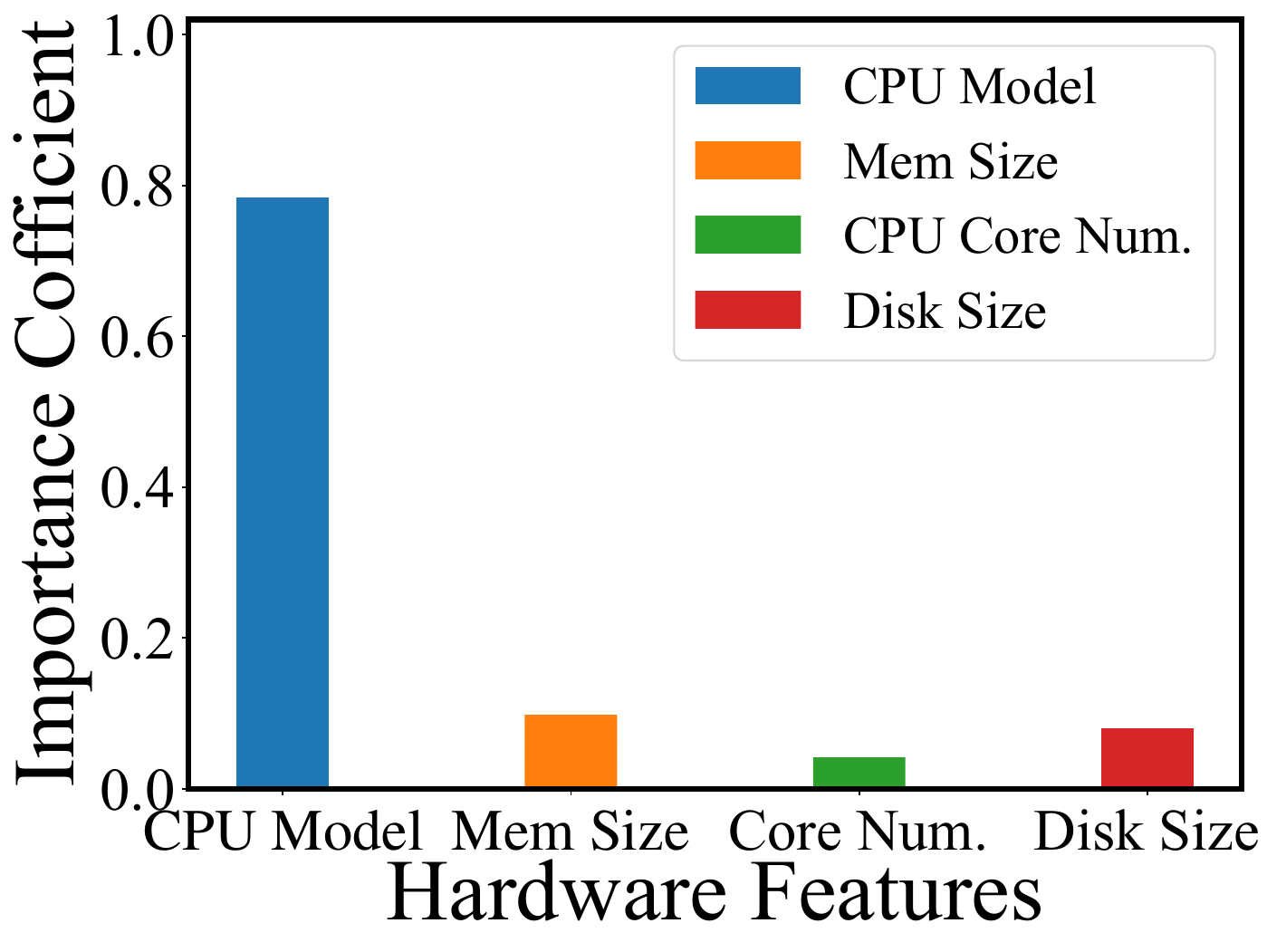}
	\caption{Importance analysis of hardware configurations}
	\label{fig:ch.3}
    \end{minipage}
	\vspace{-5mm}
\end{figure}

\section{Observations and Motivation}
\label{sec:moti}

We analyze the performance behaviors of more than 1,400 microservices, comprising over 22,000 containers within a prominent streaming video corporation. These microservices are deployed in a data center consisting of over 5,000 heterogeneous machines. Notably, 50 of these microservices are of significant scale, with more than 11,000 containers, and are responsible for critical business functionalities, such as portal services, video recommendation, streaming player services. The cumulative CPU allocation for these 50 microservices accounts for approximately 67\% of the total quota. 

\begin{table*}[tb]
\centering
\caption{Statistics of four representative microservices}
\label{tab:key_ms}
\vspace{-2mm}
{
\resizebox{0.8\textwidth}{!}{
\begin{tabular}{|c|cc|c|c|c|c|c|c|c|c|c|}
 \hline
  {\multirow{2}*{Name}}&\multicolumn{2}{c}{\multirow{2}*{{Functionality}}}&\multicolumn{3}{|c|}{Number of Containers}&\multicolumn{3}{|c|}{CPU Capacity (Cores)}&\multicolumn{3}{|c|}{Average CPU Utilization (\%)}\\
  \cline{4-12}
  &&&{Total}&{$826X$}&{$816X$}&{Total}&{$826X$}&{$816X$}&{Total}&{$826X$}&{$816X$}\\
  \cline{1-12}
   \multirow{4} *{}
   \multirow{1}*{$MS_0$} &\multicolumn{2}{c|}{Rule matching of recommendation}&\multirow{1}*{$877$}&\multirow{1}*{$603$}&\multirow{1}*{$274$}&\multirow{1}*{$3,508$}&\multirow{1}*{$2,412$}&\multirow{1}*{$1,096$}&\multirow{1}*{$30.90$}&\multirow{1}*{$29.74$}&\multirow{1}*{$33.98$}\\
   \cline{1-12} 
   
    $MS_1$&  \multicolumn{2}{c|}{ Portal requests aggregation} &$419$&$269$&$150$&$1,676$&$1,076$&$600$&$24.07$&$22.58$&$26.36$\\
    

    \cline{1-12}
    $MS_2$&  \multicolumn{2}{c|}{Advertising management}&$690$&$493$&$197$&$2,760$&$1,972$&$788$&$33.32$&$31.48$&$37.88$\\
   \cline{1-12}
    $MS_3$&\multicolumn{2}{c|}{Logging management}&$239$&$148$&$91$&$1912$&$1,184$&$728$&$25.14$&$22.66$&$28.57$\\
    \cline{1-12}
\cline{1-12}
   \hline
\end{tabular}
}
\vspace{-0.2cm}
}
\end{table*}

\begin {figure*}[tb]
    \centering
    \begin{minipage}[ht]
    {0.19\linewidth}
    \centering
    \includegraphics[scale=0.14]{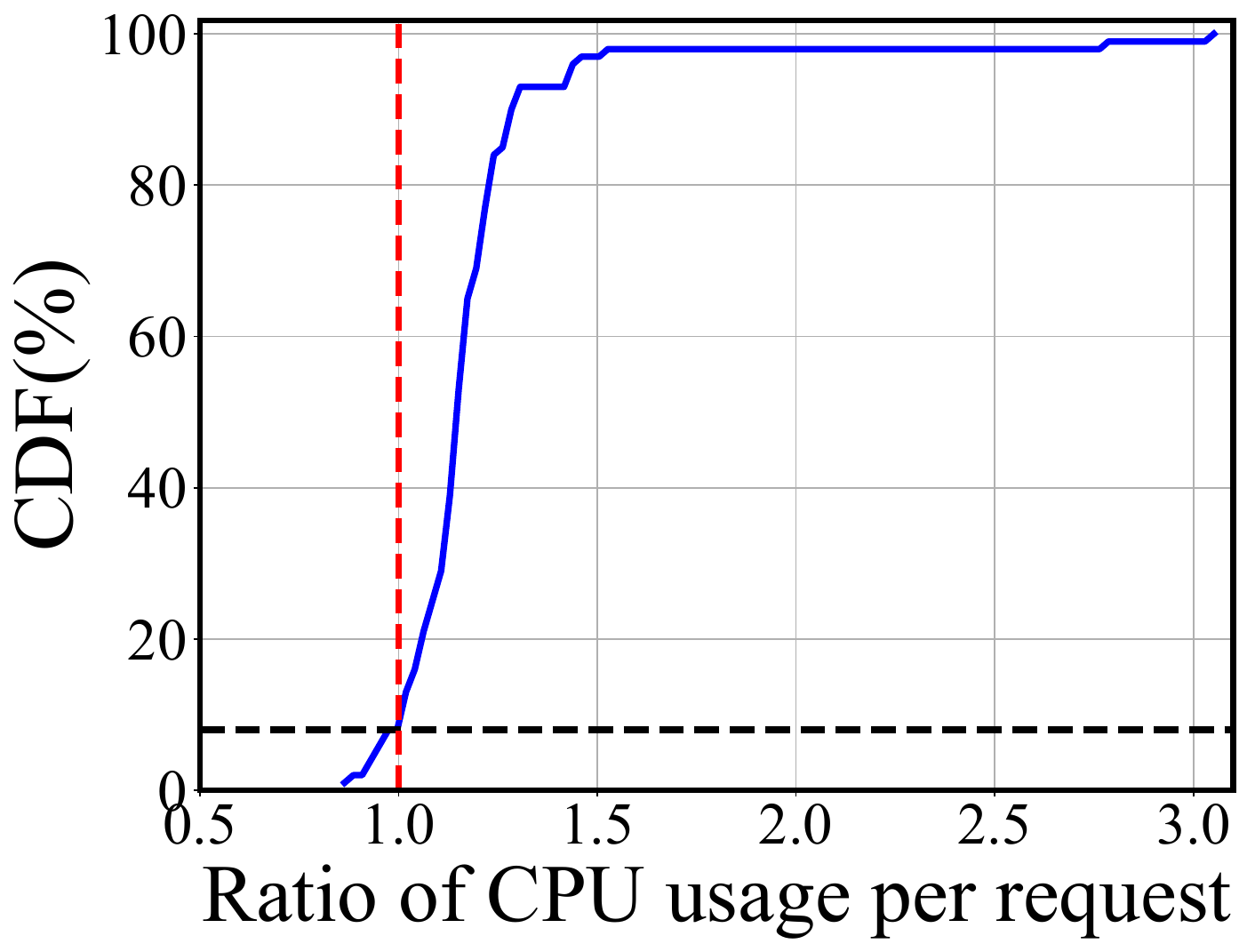}
    \caption{CDF of work efficiency ratio between $816X$ and $826X$}
  \label{fig:ch.2}
    \end{minipage} 
    \begin{minipage}[ht]{0.792\linewidth}
    \vspace{-18.5pt}
    \centering
    \subfloat[$MS_0$]{\label{fig:2.b}
\includegraphics[scale=0.146]{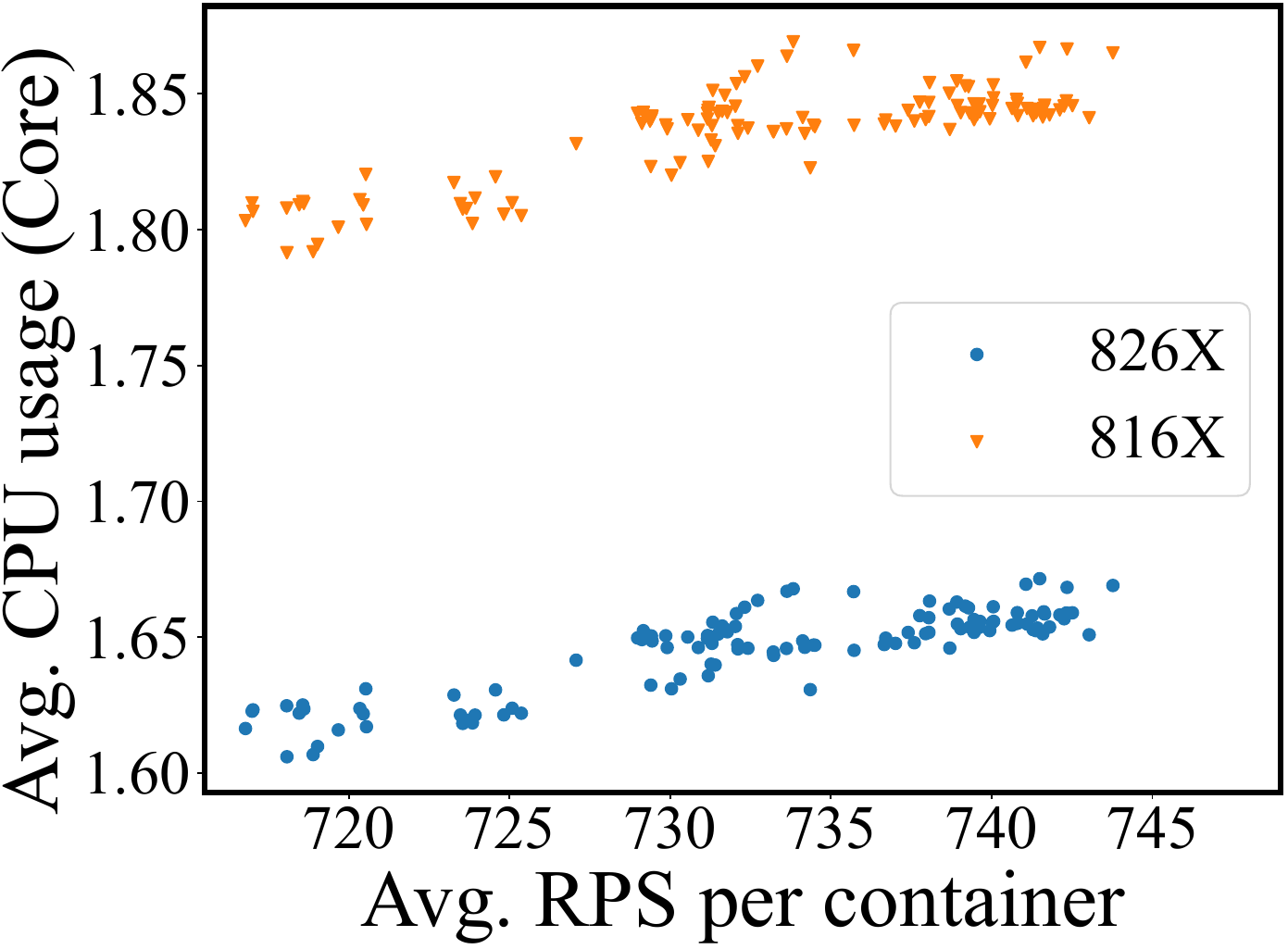}
		}
	\subfloat[$MS_1$]{\label{fig:2.c}	\includegraphics[scale=0.146]{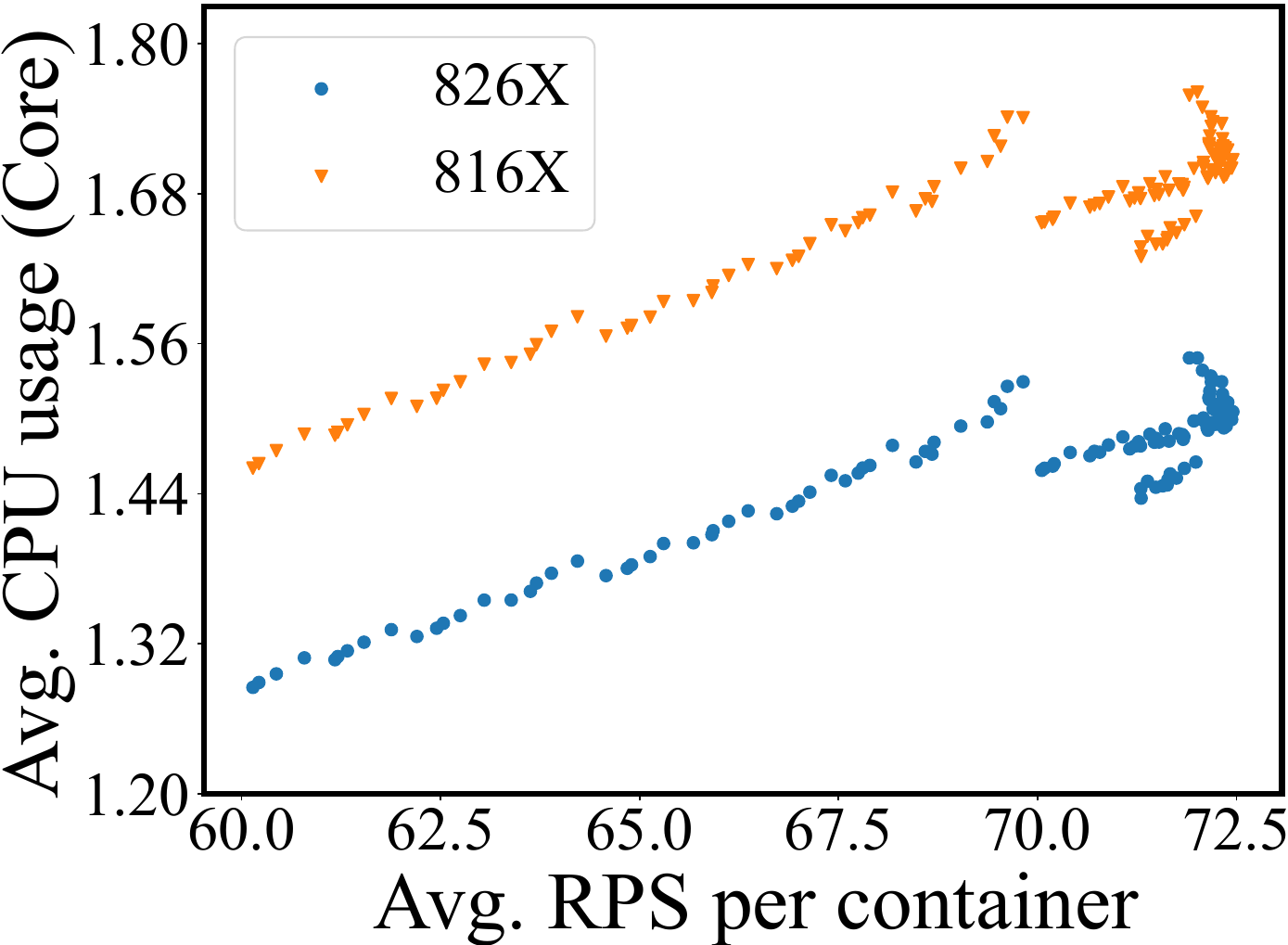}
		}
  \subfloat[$MS_2$]{\label{fig:2.d}
		\includegraphics[scale=0.146]{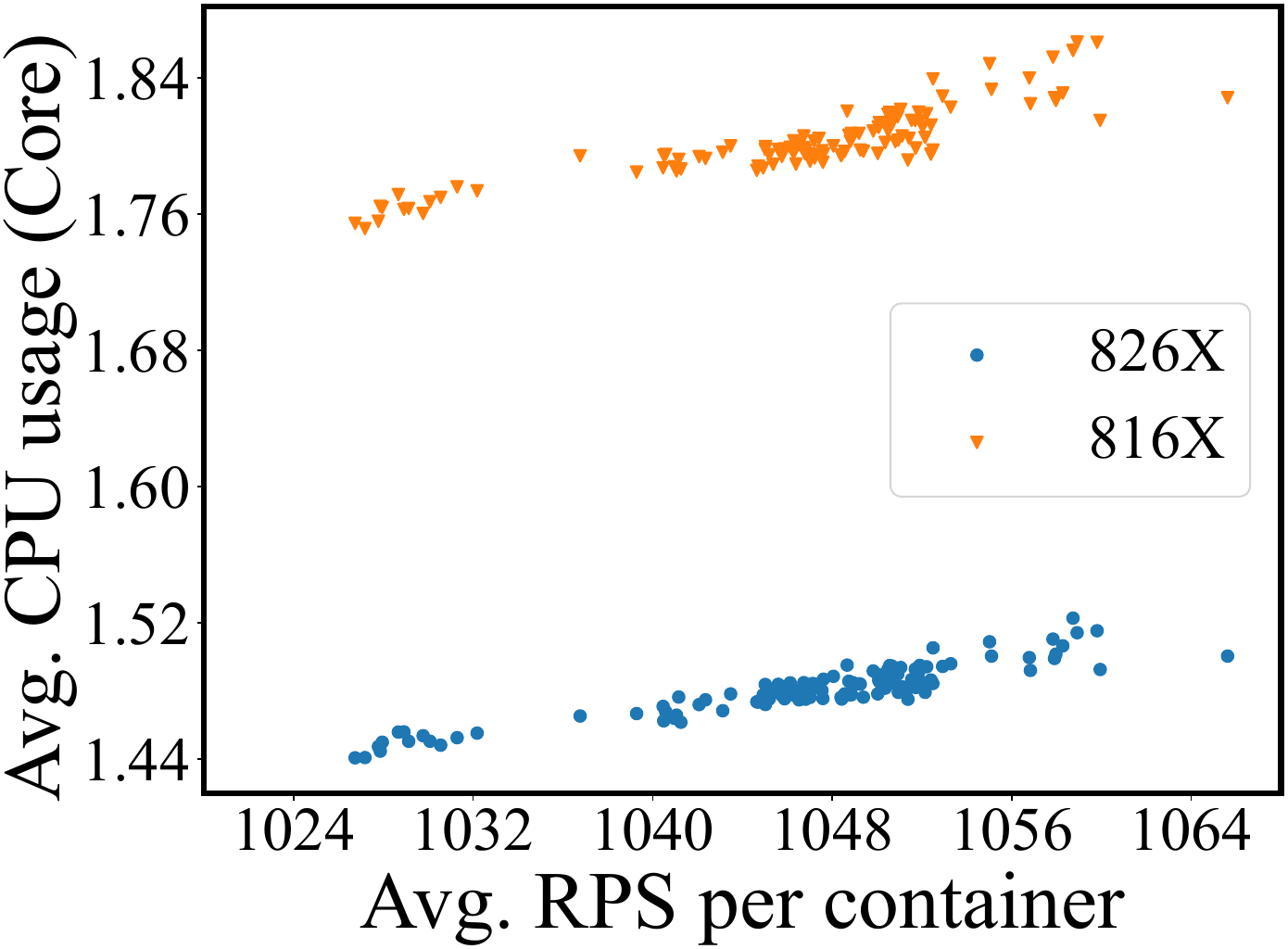}
		}
  \subfloat[$MS_3$]{\label{fig:2.e}
		\includegraphics[scale=0.146]{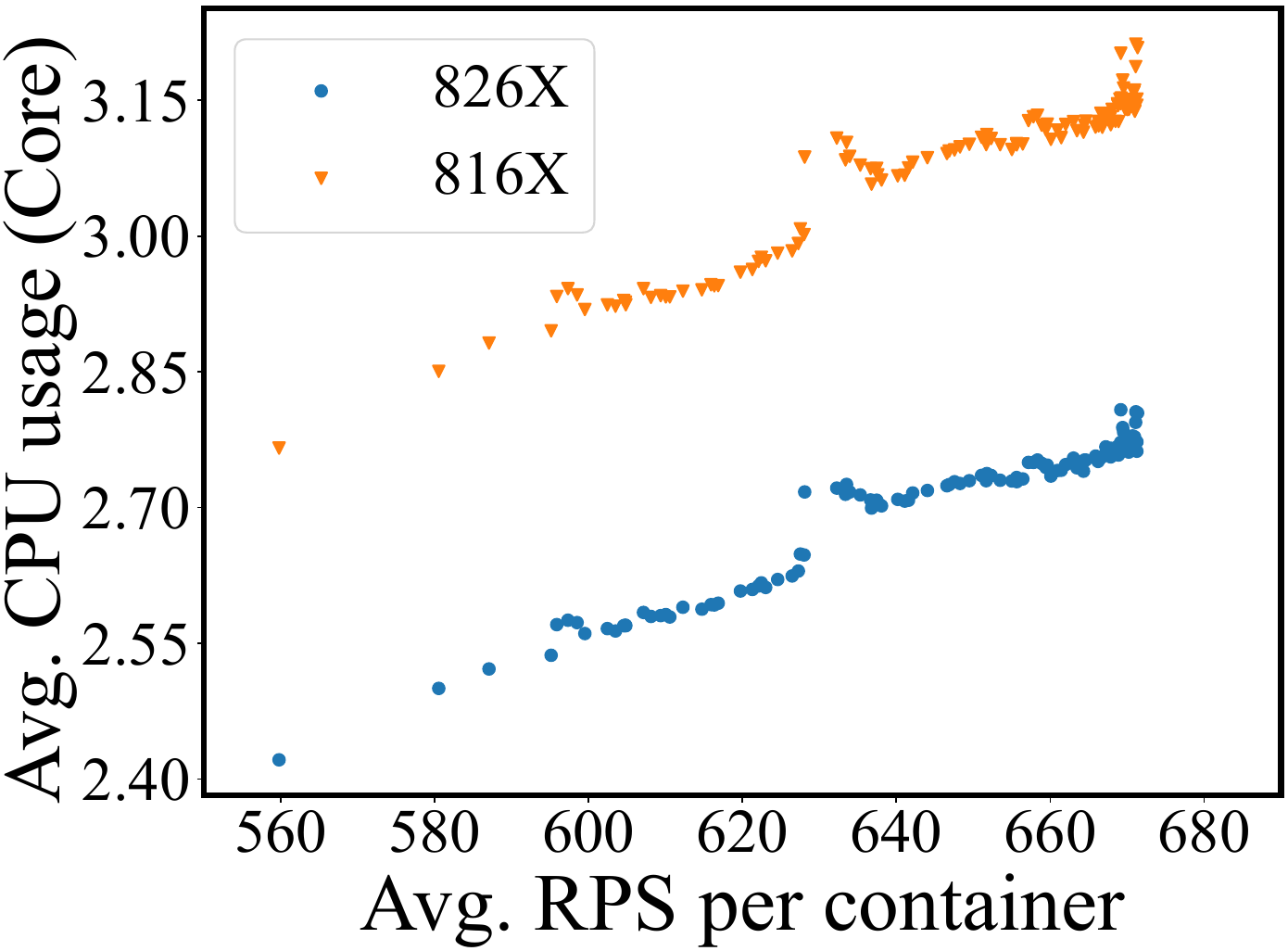}
		}
    \caption{Average container CPU usage of four microservices during peak load}	\label{fig:2}
    \end{minipage}
 	\vspace{-0.6cm}
\end{figure*}


\subsection{Overall Profiling of Microservices}
According to prior research \cite{borg,  rosen2013resource}, the relationship between CPU utilization and CPU usage\footnote{Consistent with existing literature \cite{kae-informer,proscale, rosen2013resource}, this study measures the workload of microservices in terms of requests per second (RPS). The unit of CPU usage is millicore ($mCore$), indicating the utilization of $1\text{\textperthousand}$ of a logical core per second ($1000\;mCore=1\; \text{ Core}$).} can be expressed as:
\begin{equation}
\label{eq:util_usage}
\text{CPU utilization}=\frac{\text{CPU usage}}{\text{CPU capacity}}\times 100\%.
\end{equation}
In practical scenarios, the CPU usage of microservices often exhibits a strong correlation with workload.\footnote{82.3\% of 1400 microservices have correlation coefficients exceeding 0.7.} As service workloads typically exhibit daily periodicity or long-term trends \cite{kae-informer, deep-scaling, zhou2022cushion}, CPU usage consequently fluctuates. For example, Figure \ref{fig:util_usage_rps} shows the workload and CPU usage of a microservice $MS_0$ responsible for rule-matching in video recommendation over the span of one week. The CPU utilization of $MS_0$ fluctuates in accordance with the workload.


The load balancer in production typically employs an equitable distribution of requests among the containers of each microservice, as exemplified by Google \cite{borg}, Ant\footnote{E.g., MOSN of Ant, which is accessed at: https://github.com/mosn/mosn} and Alibaba \cite{luo2021characterizing}. This approach serves to rectify workload imbalance across containers and enables an examination of how other factors contribute to the CPU usage patterns of microservices. To analyze the resource usage patterns of microservices, we select four representative ones, as outlined in Table \ref{tab:key_ms}. In terms of capacity estimation, our analysis reveals two crucial observations regarding the influence of hardware heterogeneity and service upgrades.

\subsection{Impact of Hardware Heterogeneity on Microservices}
\label{sec:char_perf_gap}
Data centers typically comprise machines with diverse hardware configurations, including architecture  \cite{sriraman2019softsku}, machine resource capacity \cite{alibaba_heter}, and others. For example, inspired by prior studies \cite{HARMONY}, we have identified 34 machine types in our data center, which exhibit heterogeneity across four crucial hardware features \footnote{{The machine network bandwidth ranges from 10 to 32 Gb/s; however, its impact on efficiency difference is negligible due to the scheduler's assurance of placing containers belonging to the same microservices on machines with the same network bandwidth configurations.}}: 1) CPU model \footnote{{In our data centers, machines with identical CPU models are equipped with corresponding memory bandwidth configurations, ensuring a consistent impact of memory bandwidth. For example, both the $826X$ and $816X$ machines are equipped with hexa-channel DDR4 memory, with memory bandwidth limits per socket of $125$ GB/s and $141$ GB/s, respectively. This discrepancy is attributed to the differing data rate of the $816X$ and $826X$ CPU models, which are $2666$ MT/s and $2993$ MT/s, respectively.}}, 2) number of CPU cores (ranging from 32 to 104), 3) memory size (ranging from 64 to 512 GB), and 4) disk size (ranging from 60 to 3,880 GB).


We employ XGBoost \cite{xgboost} to examine the relative significance of the four hardware features on the work efficiency ratio of various machine types.\footnote{For each microservice, we collect CPU usage of containers on each machine type. The load balancer guarantees the equitable workload intensity among them at each time-point. XGBoost uses the four features to predict the work efficiency ratio among machine types, and the relative importance of each feature is quantified as the average gain in prediction performance achieved across all gradient boosted trees, as elaborated in \cite{elith2008working}.} As depicted in Figure \ref{fig:ch.3}, the average relative importance of the CPU model can reach 0.783 for the 50 large-scale microservices, indicating that diverse CPU models can be considered the predominant factors contributing to the disparity in work efficiency among different hardware configurations. Therefore, we concentrate on the two primary CPU models in our data center, which collectively account for approximately 95\% of all machines: Intel Xeon Platinum {\bf 826X} (Cascade Lake model) and Intel Xeon Platinum {\bf 816X} (Skylake model). These two-generation CPU models possess a range of distinct micro-architectural characteristics. For example, compared with $816X$, $826X$ exhibits an increase in turbo frequency by approximately $100\sim300$MHz ($2.5\%\sim7.5\%$) and a 10\% enhancement in data access rate (from 2666 MT/s to 2933 MT/s).

CPU utilization is a crucial metric for assessing resource efficiency and service performance in the context of stateless and lightweight microservices \cite{deep-scaling, rzadca2020autopilot}. Consequently, average CPU usage serves as an indicator of the work efficiency of microservices deployed on heterogeneous machines. In this regard, a lower CPU usage signifies enhanced work efficiency when subjected to the same workload \cite{yi2020cpi}. 


We investigate the performance disparity among 50 large-scale microservices deployed on the two CPU models. Figure \ref{fig:ch.2} shows the cumulative distribution function (CDF) of the ratio of average CPU usage per request between the $816X$ and $826X$ machines. Specifically, approximately 84\% of these microservices on the $816X$ machines exhibit a CPU consumption that is 0.5\%-25\% higher than that of the $826X$ machines. Conversely, around 8\% of these microservices on the $816X$ machines demonstrate an enhancement in resource efficiency, ranging from 2.6\% to 13.5\%. Additionally, we analyze the performance of four representative microservices, namely $MS_0$ to $MS_3$, under peak load conditions. This analysis is critical for resource adjustment to accommodate maximum workloads. As shown in Figure \ref{fig:2}, the containers on $826X$ machines are capable of reducing the average peak CPU usage by about 12.3\% ($MS_0$) to 22.9\% ($MS_2$) when compared to $816X$ machines.

{\bf Observation 1}: 
Variations in CPU characteristics can result in notable fluctuations in the CPU utilization of microservices. Additionally, microservices exhibit varying levels of sensitivity towards CPU heterogeneity.

{\bf Implication:} To ensure precise resource allocation for microservices under a dynamic workload, it is imperative to assess the resource efficiency of individual microservices on distinct CPU models. 

\subsection{Impact of Version Upgrades on Microservices}
\label{sec:upgrade}
\begin{figure*}[t]
     \centering
     \subfloat[{The traces of the number of \\upgraded containers}]{\label{fig:up.trace}
		\centering \begin{minipage}[t]{0.24\linewidth}
		\setlength{\abovecaptionskip}{-0.03cm}
		 \centering
\includegraphics[scale=0.182] {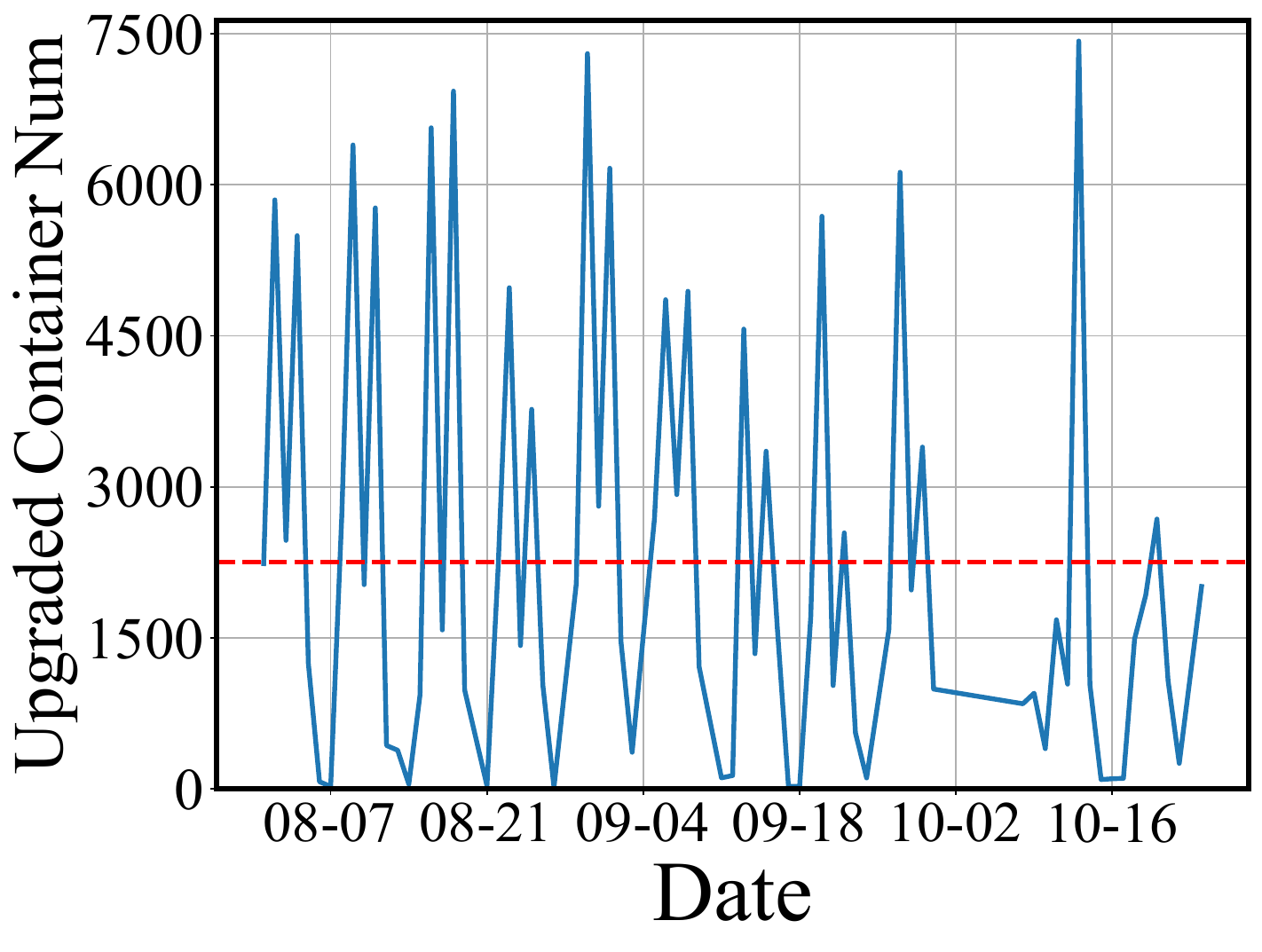}
		\end{minipage}
		}
	\subfloat[CDF of upgrade interval of\\50 large-scale microservices]{\label{fig:up.1}
		\centering \begin{minipage}[t]{0.24\linewidth}
		\setlength{\abovecaptionskip}{-0.03cm}
		 \centering
\includegraphics[scale=0.18] {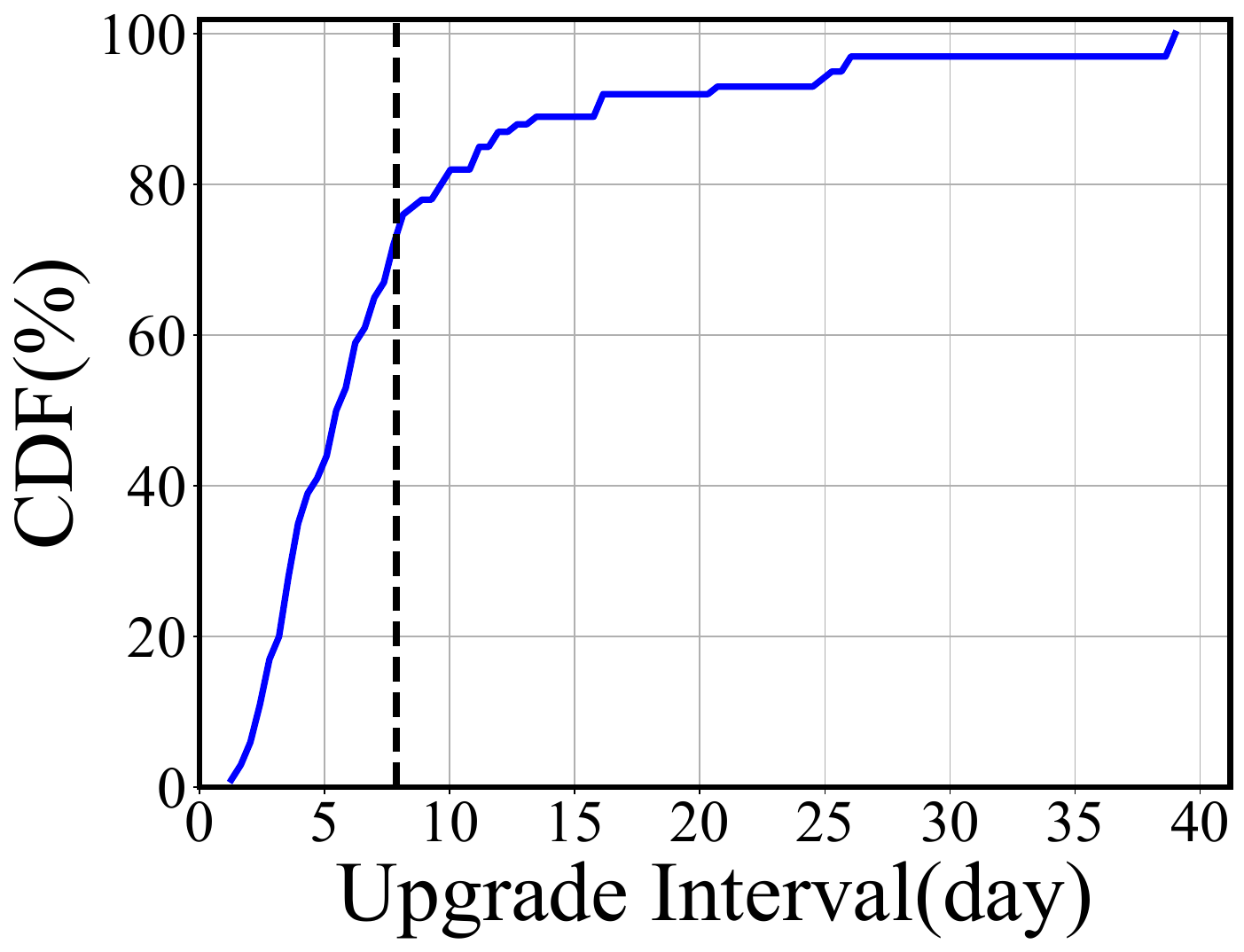}
		\end{minipage}
		}
	\subfloat[CDF of upgrade overhead of 50 large-scale microservices]{\label{fig:up.2}
		\begin{minipage}[t]{0.24\linewidth}	\setlength{\abovecaptionskip}{-0.03cm}
		\centering
\includegraphics[scale=0.18]{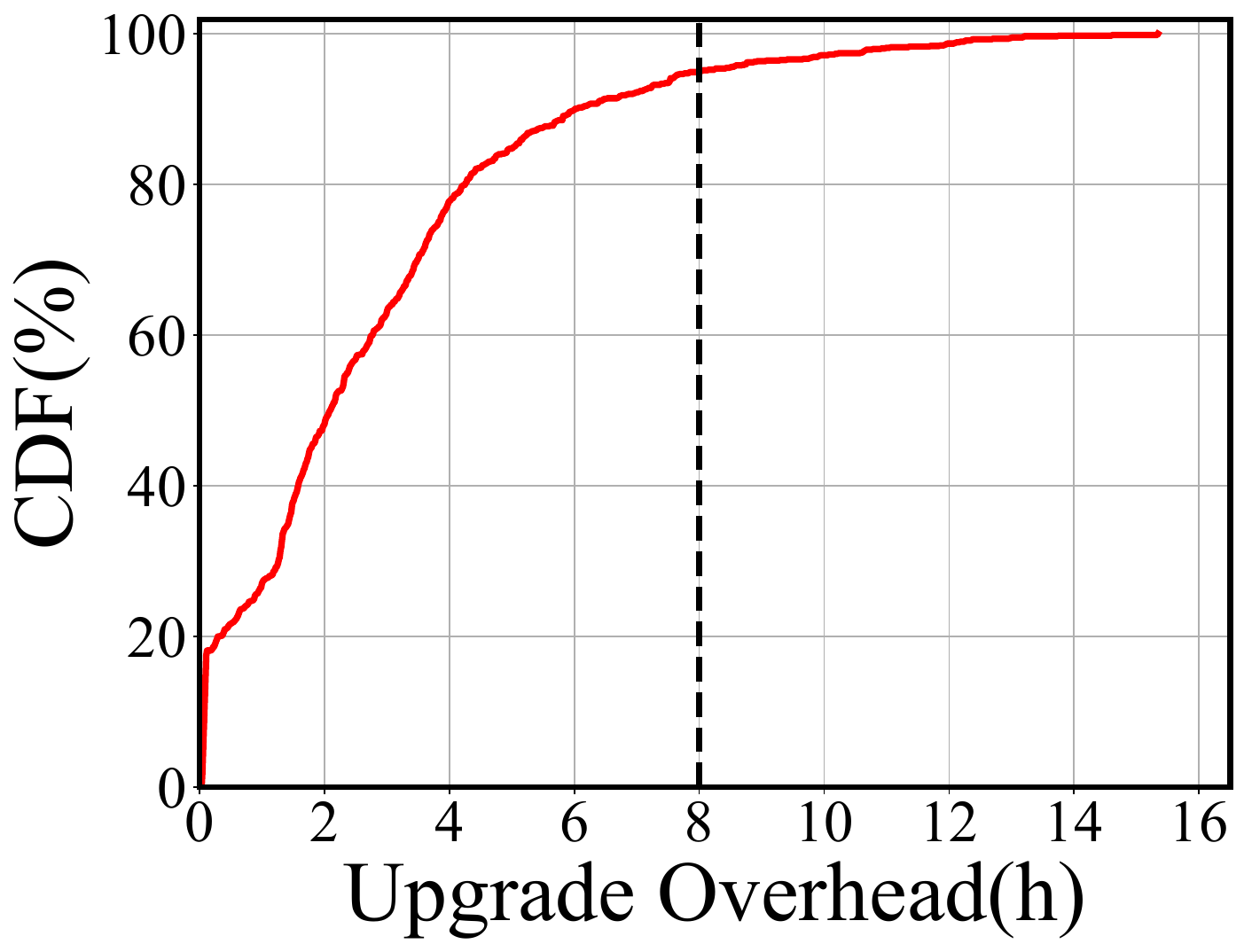}
		\end{minipage}
		}
  \subfloat[CDF of change rate of CPU usage per request due to upgrades]{\label{fig:up.total}
		\begin{minipage}[t]{0.24\linewidth}	\setlength{\abovecaptionskip}{-0.03cm}
		\centering
\includegraphics[scale=0.178]{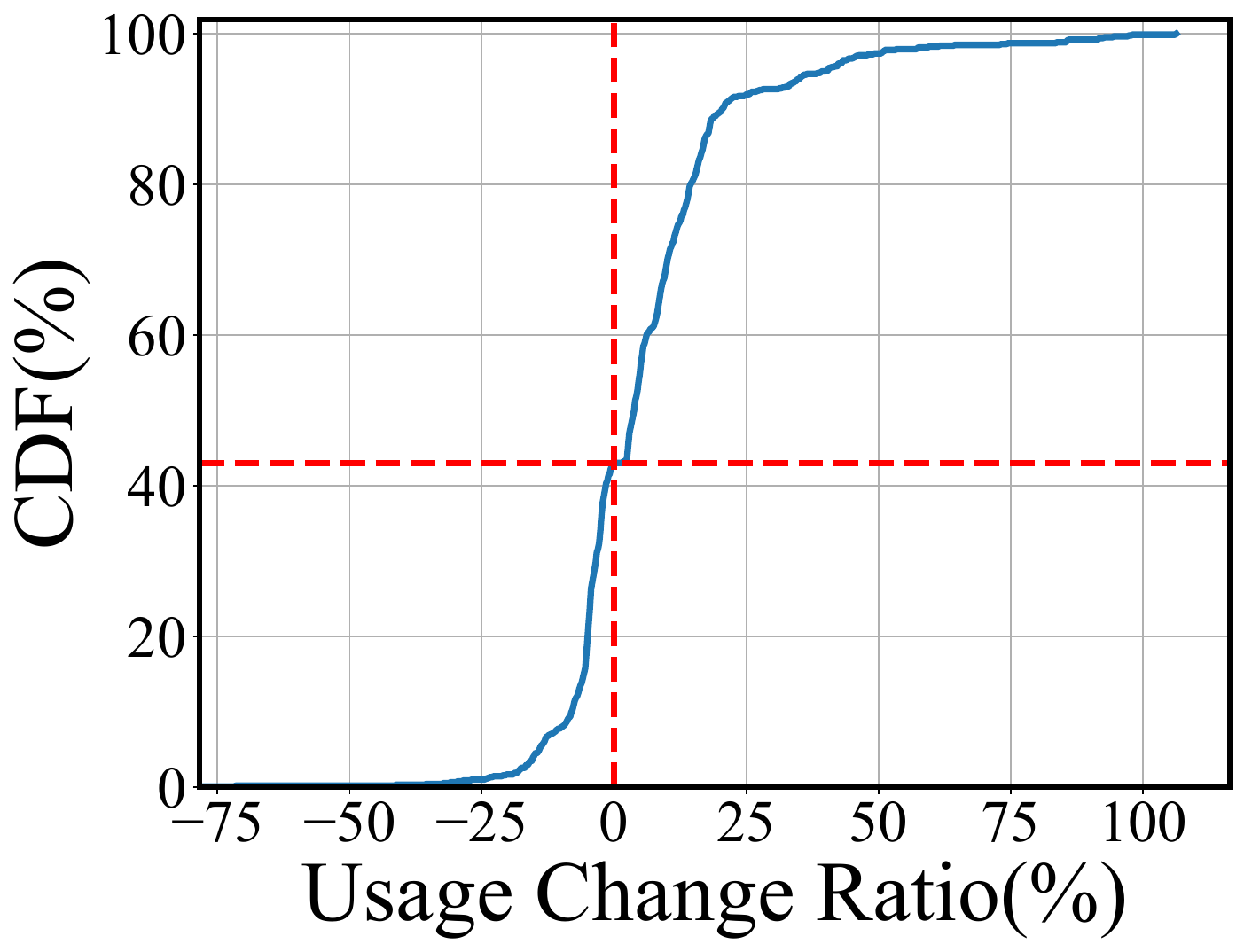}
		\end{minipage}
		} 
    \caption{The unpredictability of version upgrades}
    \label{fig:ups}
 	\vspace{-0.4cm}
\end{figure*}

\begin{figure*}[t]
     \centering  
	\subfloat[Case 1 ($MS_0$)]{\label{fig:3.b}		\centering \begin{minipage}[t]{0.24\linewidth}
		\setlength{\abovecaptionskip}{-0.02cm}
		 \centering
  \includegraphics[width=\textwidth] {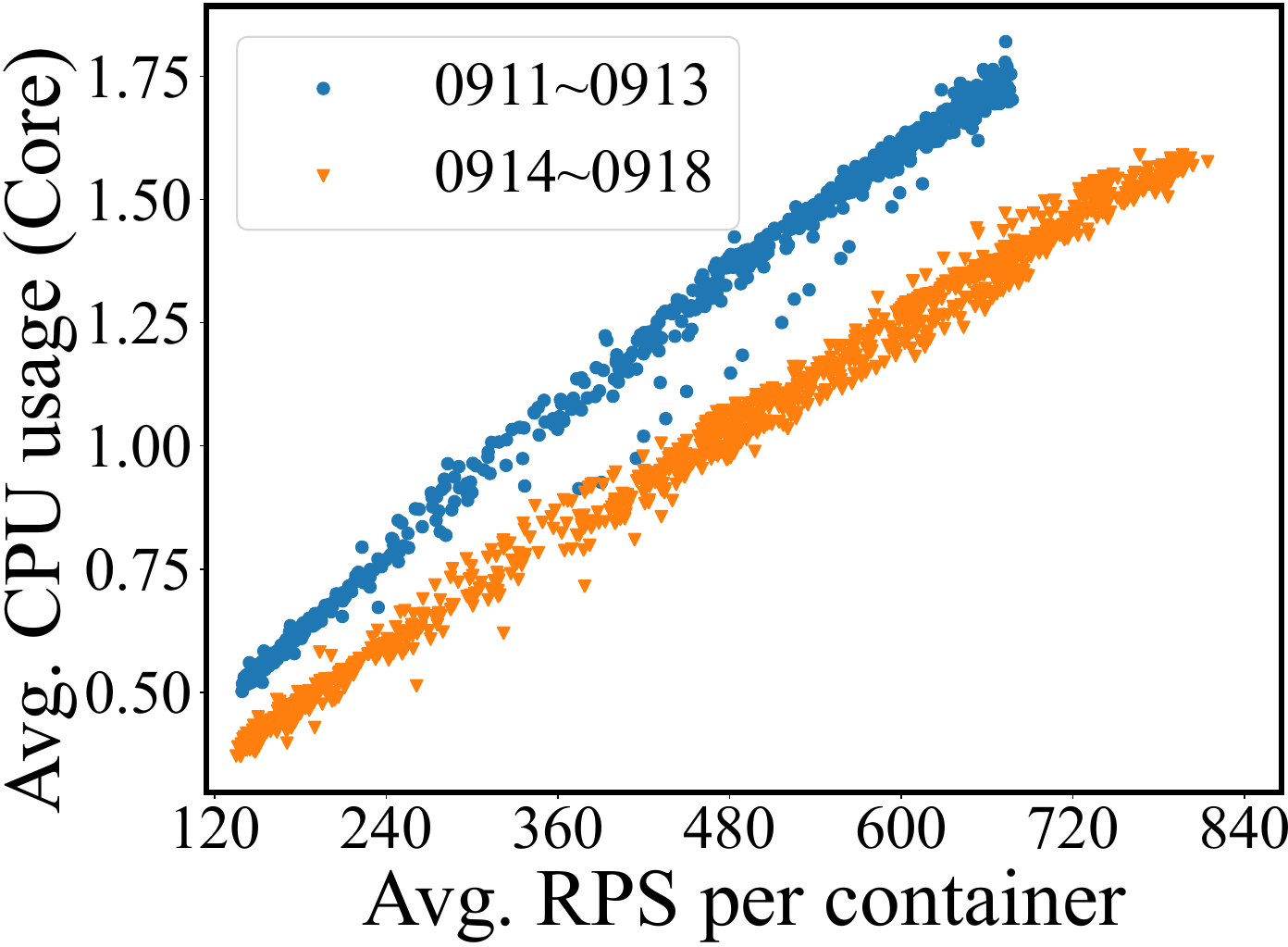}
		\end{minipage}
		}
	\subfloat[Case 2 ($MS_1$)]{\label{fig:3.c}
		\begin{minipage}[t]{0.24\linewidth}
		\setlength{\abovecaptionskip}{-0.02cm}
		\centering
    \includegraphics[width=\textwidth]{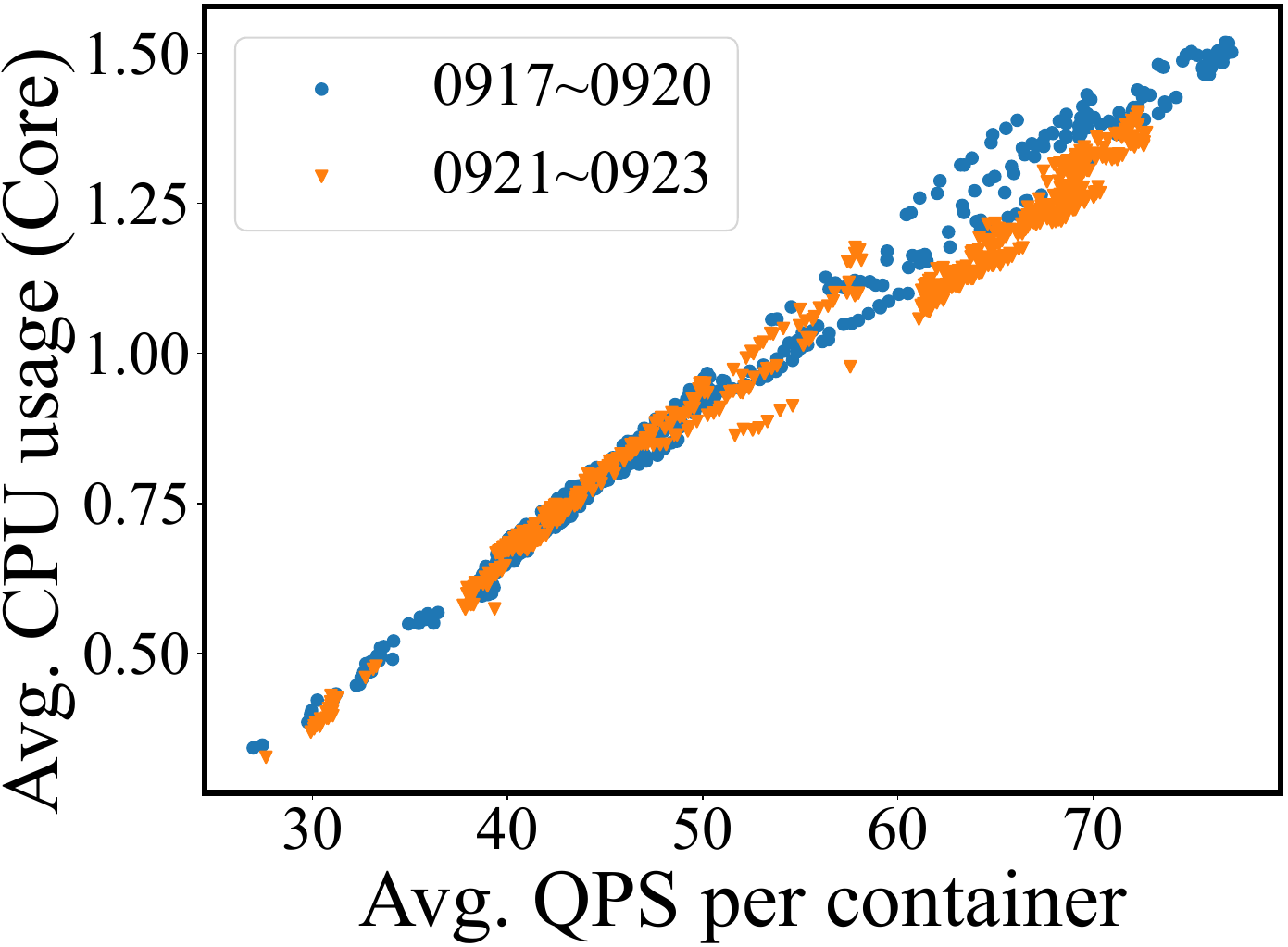}
		\end{minipage}
		}
  \subfloat[Case 3 ($MS_2$)]{\label{fig:3.d}
		\begin{minipage}[t]{0.24\linewidth}
		\setlength{\abovecaptionskip}{-0.02cm}
		\centering
    \includegraphics[width=\textwidth]{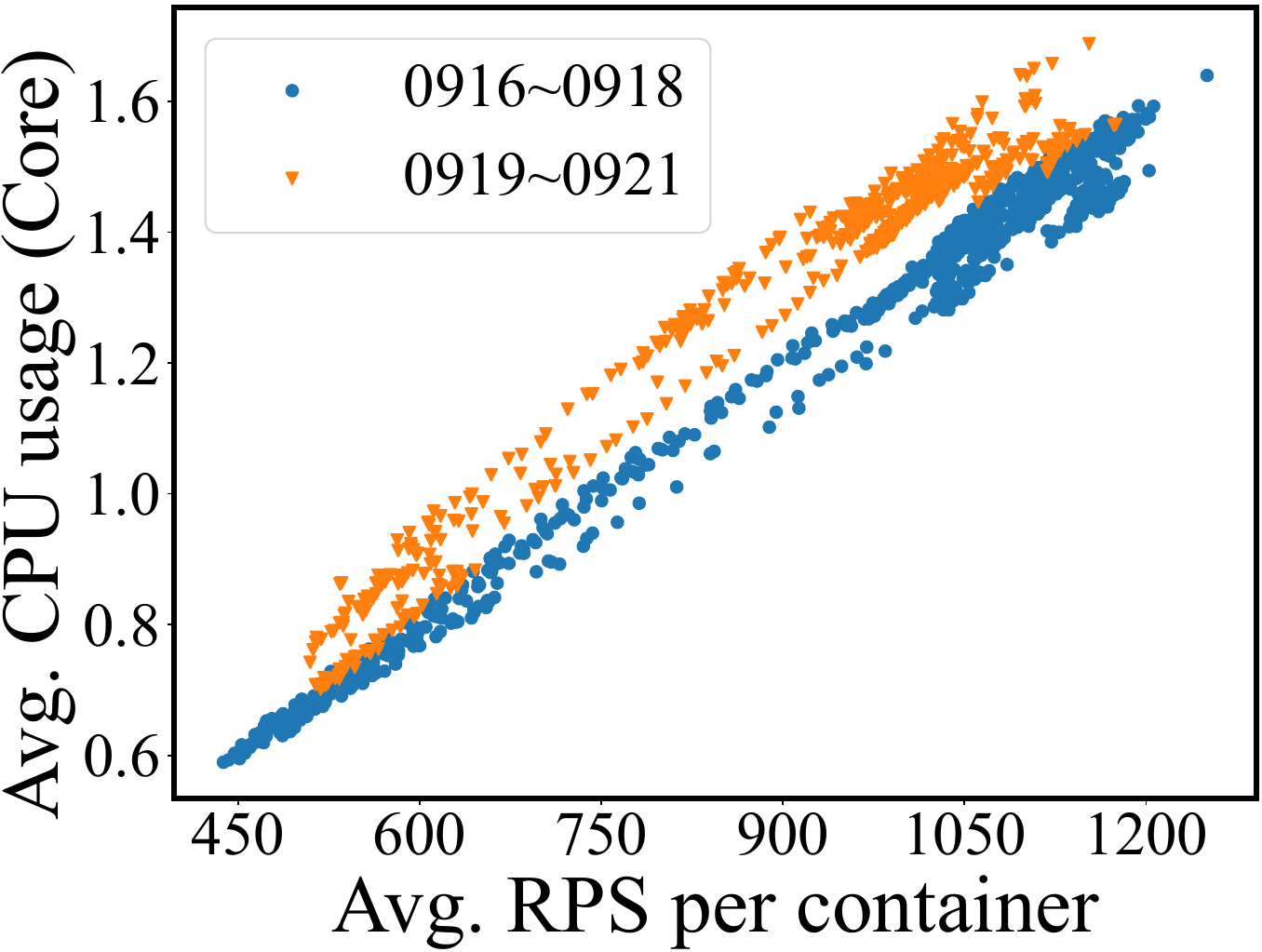}
		\end{minipage}
		}
  \subfloat[Case 4 ($MS_3$)]{\label{fig:3.e}
		\begin{minipage}[t]{0.24\linewidth}
		\setlength{\abovecaptionskip}{-0.02cm}
		\centering
 \includegraphics[width=\textwidth]{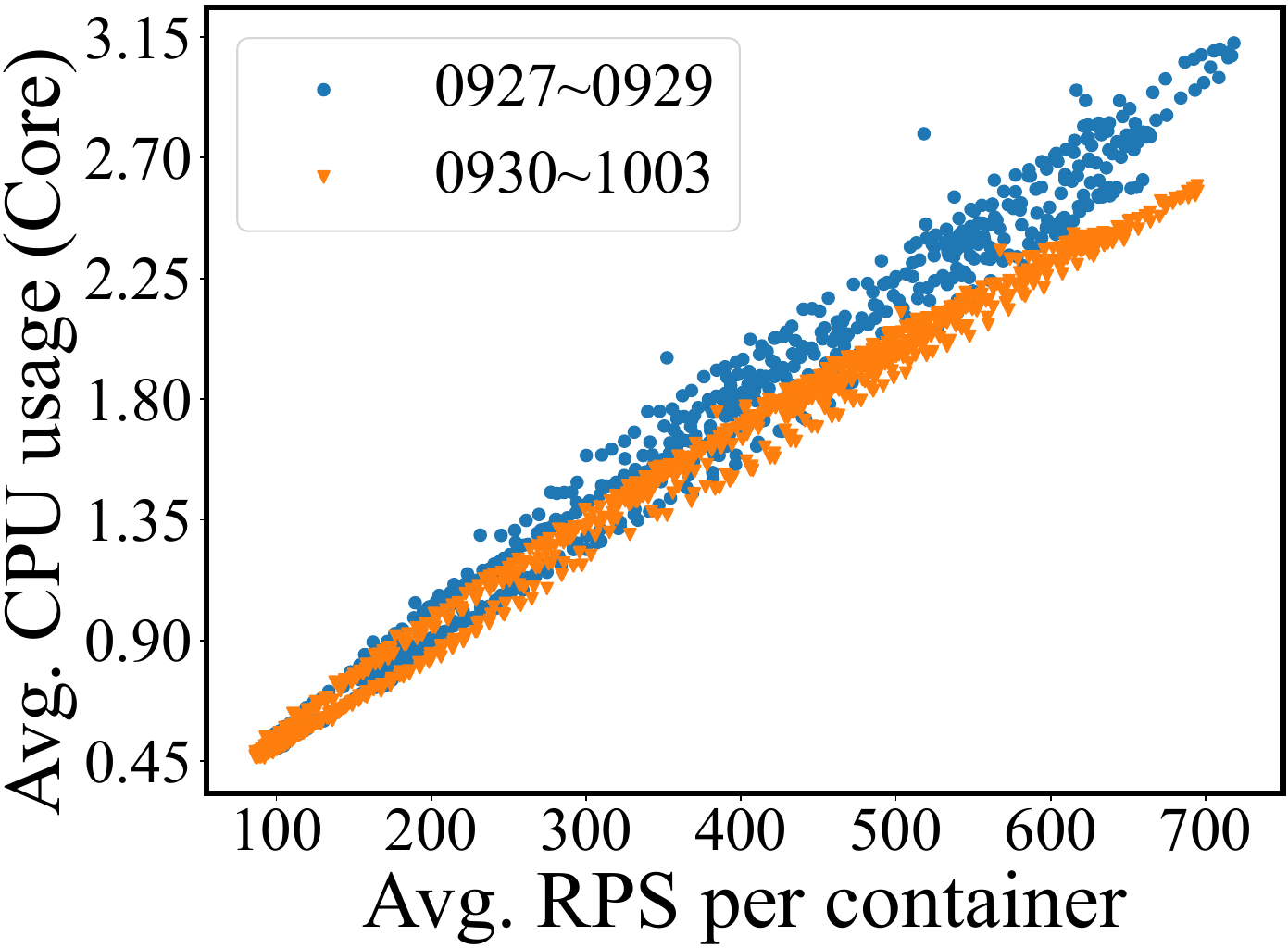}
		\end{minipage}
		}
	\vspace{-1mm}
    \caption{The four cases of CPU usage before and after the upgrade}
    \label{fig:3}
 	\vspace{-0.4cm}
\end{figure*}

Supported by continuous integration and continuous delivery (CI/CD) tools \cite{8776985}, as well as container technology, the implementation of version upgrades has become a regular and frequent practice in microservices, aimed at meeting ever-evolving business requirements. For instance, within the timeframe of Aug. to Oct. 2022, a total of 922 version upgrades were performed for the 50 microservices. Furthermore, on a daily basis, an average of over 2,000 containers undergo upgrades, as shown in Figure \ref{fig:up.trace}. 

These upgraded microservices exhibit two distinct characteristics that are driven by business needs. Firstly, the frequency of upgrades is frequent but difficult to accurately anticipate. This variability is illustrated in Figure \ref{fig:up.1} and Figure \ref{fig:up.2}, where the average interval between upgrades ranges from a few days to several dozen days, with 95\% of upgrades being completed within 8 hours. Secondly, each upgrade has a notably uncertain impact on the CPU usage of the respective microservice. This uncertainty is depicted in Figure \ref{fig:up.total}, which demonstrates that the average change rate of CPU usage per request for microservices after upgrades varies significantly, $-77.1\%\sim106.1\%$. Additionally, approximately 64\% of upgrades exhibit substantial alternations in CPU usage ($\big|\text{change rate}\big|>5\%$). Furthermore, when subjected to the same workload, about 43\% of upgrades result in a reduction in CPU usage, while the remaining 57\% lead to an increase. To elucidate this phenomenon, we conduct an analysis of the resource utilization patterns of the four representative microservices, namely $MS_0$ to $MS_3$.

{\bf Case 1: Upgrades to optimize performance.} To minimize the computational overhead associated with rule matching for video recommendations in $MS_0$, the engineers submitted optimized code for $MS_0$ on Sep. 13, 2022. As shown in Figure \ref{fig:3.b}, this upgrade results in a reduction of approximately 21.6\% in average CPU usage.

{\bf Case 2: Upgrades to fix bugs.} On Sep. 20, 2022, a bug was identified in the advertisement filtering algorithm of $MS_1$. Upgrades were implemented by the developers to rectify the issue, leading to an average CPU usage reduction of around 2.2\% for $MS_1$, as shown in Figure \ref{fig:3.c}.

{\bf Case 3: Upgrades to implement new features.} On Sep. 19, 2022, developers made modifications to the payment models in $MS_2$. As illustrated in Figure \ref{fig:3.d}, following this upgrade, the average CPU usage of $MS_2$ experiences an increase of about 11.1\%.

{\bf Case 4: Upgrades to incorporate new functionalities.}
On Sep. 30, 2022, new functionalities were integrated into $MS_3$ to capture payment logs for various types of mobile devices. This upgrade leads to a reduction in the average CPU usage of $MS_3$ by around 5.1\%, as depicted in Figure \ref{fig:3.e}.

{\bf Observation 2:} The influence of version upgrades on the performance of microservices is highly uncertain. Furthermore, the extent of CPU usage alternation caused by upgrades varies significantly.

{\bf Implication:} The significance of frequent version upgrades cannot be disregarded. The timely capture of changes in microservice CPU usage caused by version upgrades is crucial for the accurate estimation of resource capacity. For instance, following the upgrades in Case 3, capacity planning that neglects upgrades leads to a peak CPU utilization of approximately 70\% in the containers of $MS_2$, resulting in a rise in tail latency by around 21.9\% and causing notable performance anomalies.

\subsection{Motivation}
The key to implement resource auto-scaling for microservices lies in accurately modeling their performance patterns. As indicated in Observations 1 and 2, the CPU usage pattern of microservices is greatly influenced by hardware heterogeneity and version upgrades. Therefore, it is imperative to explicitly take these factors into account, which serves as the driving force behind the design of Humas.

\section{{System Design}}
\label{sec:mehtod}
\subsection{Overview}
The essence of Humas lies in capturing variations in microservice CPU usage patterns, as this forms the foundation for the accurate estimation of resource capacity. By proactively accounting for the effects of hardware heterogeneity and version upgrades on microservice CPU usage patterns, Humas facilitates prompt and precise capacity auto-scaling, thereby ensuring consistent performance. 

\begin{figure}[t]
	\centering	\includegraphics[width=0.472\textwidth]{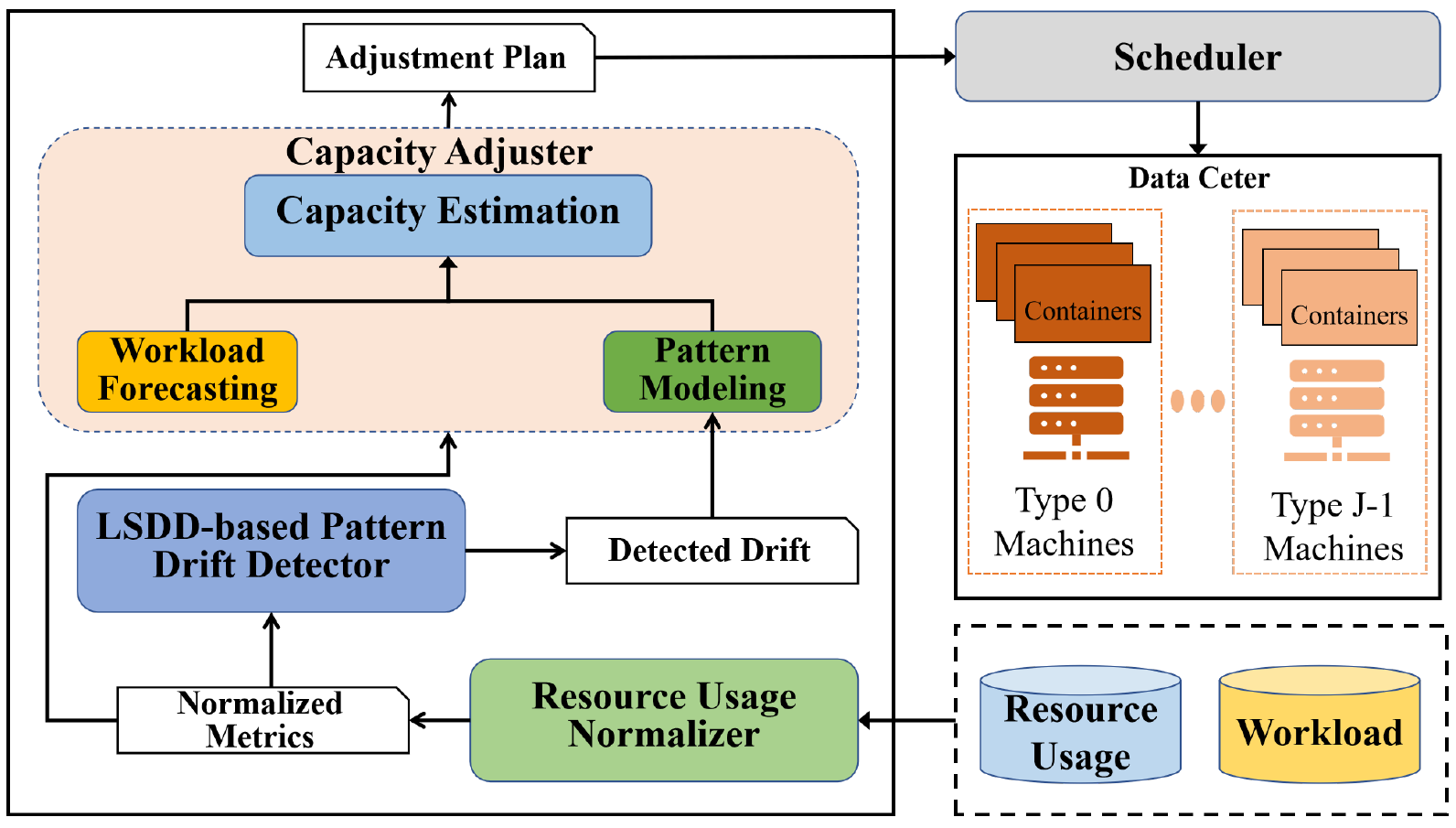}
	\caption{The framework of Humas}
	\label{fig:framework}
 	\vspace{-0.4cm}
\end{figure}

Figure \ref{fig:framework} illustrates the framework of Humas, comprising three modules. Firstly, the normalizer unifies the resource utilization efficiency of the microservices deployed on heterogeneous machines. Secondly, the drift detector, employing the LSDD indicator, is devised to detect alternations in patterns resulting from version upgrades. Lastly, the adjuster is tasked with determining the capacity adjustment plan for each microservice in the dynamic environment.

The workflow of Humas is as follows. Initially, Humas retrieves the performance metrics at the container level for each microservice from two perspectives: 1) the workload expressed in terms of RPS, and 2) CPU usage. Subsequently, the CPU usage of containers deployed on different types of machines is normalized to unify their efficiency. Next, Humas continuously detects changes in CPU usage patterns through an online hypothesis testing method. Upon identifying a drift, the usage pattern undergoes re-training using the most recent performance data. The adjuster estimates the future resource usage based on the updated pattern and the predicted workload and generates a capacity adjustment plan. Finally, Humas engages with the scheduler to fulfill auto-scaling. Humas is designed to be non-intrusive, allowing it to seamlessly integrate with any scheduler.

\subsection{Resource Usage Normalizer}
\label{sec:heter_stander}

The core of the resource normalizer lies in devising an efficient approach for gauging the disparity in performance across heterogeneous machines when dealing with different microservices that possess dynamic workloads. To achieve this, two factors are taken into consideration. Firstly, as explicated in Observation 1, microservices execute diverse business logics, thereby exhibiting varying sensitivities towards hardware heterogeneity. Hence, the method employed to evaluate the discrepancy in efficiency should be applicable across a wide range of microservices. Secondly, the metric used to assess work efficiency should effectively adapt to the dynamic workload prevalent in the production environment.

\paragraph{Measurement of Resource Efficiency Difference} 
While the CPI metric is commonly used to assess micro-architectural performance \cite{yi2020cpi}, it is unsuitable for quantifying the work efficiency at the business level when confronted with dynamic workloads. This limitation arises from the fact that the CPI of microservices is contingent upon workload intensity. Inspired by Intel's research \footnote{More details are elaborated at: {https://www.datacenterdynamics.com/en/\\whitepapers/alibaba-realizing-computing-power-in-hyper-scale-cloud-clusters/}}, we employ a metric  known as resource usage effectiveness (RUE) \cite{yi2020cpi} to gauge the work efficiency of microservices at the business level for two main reasons. Firstly, RUE signifies the average resource consumption per request in microservices and is directly modeled as the ratio between resource usage and workload, making it independent of workload intensity \cite{yi2020cpi}. Secondly, RUE is a versatile metric capable of measuring efficiency across a variety of microservices. This is due to the fact that business-level requests can encompass a multitude of types \cite{deep-scaling}, such as RPC or HTTP requests originating from upstream services or users \cite{Meta,sinan,proscale}, access requests directed towards cloud databases \cite{P-Store}, and message publications/subscriptions to/from middleware brokers \cite{ANTONIC2016607}.

Let us consider a scenario where there are $J$ CPU models. For any given microservice $\tau$, we denote the number of containers running on machines of type $j$ at each timestamp $t$ as $n^j_t$.
Subsequently, we aggregate the following two metrics among the $n^{j}_t$ containers of $\tau$: 1) $\overline{RPS}^{j}_{t}$, which represents the average workload of each container; and 2) $\overline{CU}^{j}_{t}$, which denotes the average CPU usage of each container. Ultimately, the RUE value for $\tau$ on machines of type $j$ at time $t$ is computed as:
\begin{equation}
\label{eq:rue_define}
RUE^{j}_{t}=\frac{\overline{CU}^{j}_{t}}{
\overline{RPS}^{j}_{t}
}
\end{equation}

An increase in RUE signifies a decrease in resource efficiency.

Then, we designate the CPU model that encompasses the highest proportion of machines as the standard type. Based on $RUE$, we define the resource efficiency difference, denoted as $RED^{j}$, between $j$-type machines and standard machines as:
\begin{equation} 
    \label{eq:red}
    RED^{j} = \mathbb{E}\left(\frac{RUE^{j}_{t}}{RUE^{std}_{t}}\right),
\end{equation}
where $\mathbb{E}()$ represents the expectation function. A value of $RED^{j} > 1$ indicates that running $\tau$ on machines of type $j$ diminishes resource efficiency in comparison to the standard machines, and vice versa.

\paragraph{Resource Usage Normalization}  
Based on $RED$, we normalize the average resource usage of containers deployed on machines of type $j$ by:
\begin{equation} 
\label{eq:stand}
\overline{CU}^{j\prime}_t = \frac{\overline{CU}^{j}_t}{RED^{j}}
\end{equation} 

Following the normalization step, we aggregate the two metrics of all containers belonging to $\tau$ on various types of machines at each time $t$. These metrics include: 1) $\overline{RPS}_t$, the average workload of each container; and 2) $\overline{CU}^{\prime}_t$, the average normalized CPU usage of each container.

\subsection{Resource Usage Pattern Drift Detector}
\label{sec:detector}
Motivated by the insights gained from Section \ref{sec:upgrade}, microservices undergo frequent upgrades to align with evolving business requirements, resulting in changes in CPU usage patterns, commonly referred to as pattern drifts. To ensure precise capacity estimation based on the most recent CPU usage patterns, we propose an online detector for identifying CPU usage pattern drifts. The detector operates in two stages. Firstly, we establish a mathematical formulation for characterizing pattern drift phenomena and introduce a sliding window approach to quantify these changes. Secondly, we present a mechanism for detecting pattern drifts, allowing for the effective monitoring and timely detection of pattern variations caused by version upgrades.

\begin{figure}[t]
	\centering	\includegraphics[width=0.38\textwidth]{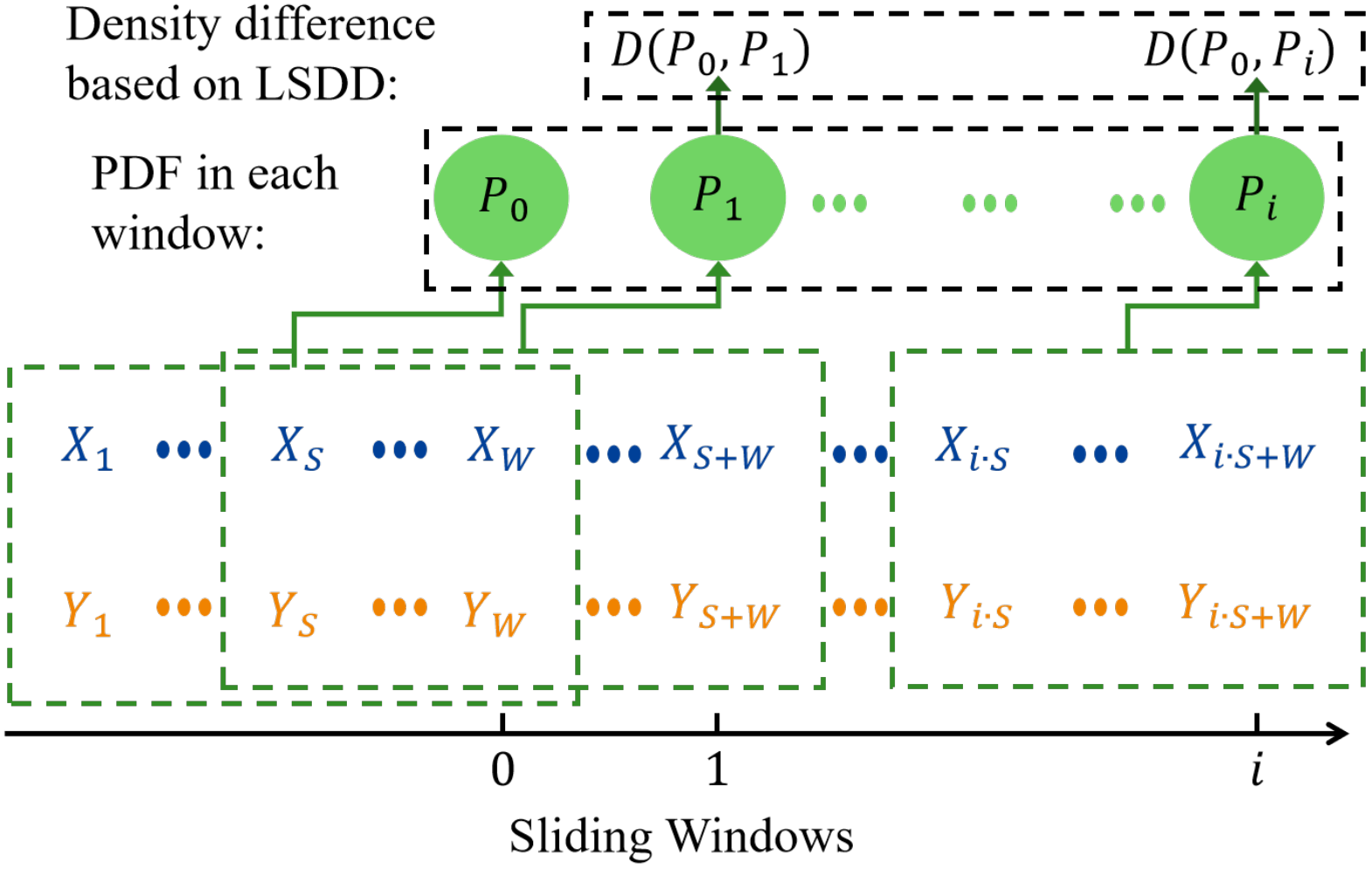}
 \vspace{-2mm}
    \caption{Estimating the density difference over sliding windows}
  \label{fig:slid.1}
 	\vspace{-4mm}
\end{figure}


\paragraph{Formulation of Pattern Drifts}
Usage patterns essentially represent the probability density functions (PDFs) of the joint distribution for CPU usage and workload \cite{Meta,deep-scaling,sinan}. Thus, pattern drifts can be identified through the changes observed in these distributions over time \cite{lu2014concept}. To capture and quantify pattern drifts accurately, we employ a sliding window approach that effectively measures the difference in probability density between adjacent windows.

For any given microservice $\tau$, we calculate the sequences of total workload and normalized CPU usage, denoted as $\mathbb{X}=\{X_{0},X_{1},...\}$ and $\mathbb{Y}=\{Y_{0},Y_1,...\}$ respectively, using the following equations:
\begin{equation}
\label{eq:total_data}
\begin{aligned}
X_t &= \overline{RPS}_t \times n_t \\
Y_t &= \overline{CU}^{\prime}_t \times n_t \\
\end{aligned}
\end{equation}
Here, $t$ represents the sampling timestamp, and $n_t$ denotes the total number of containers associated with microservice $\tau$ at time $t$. For simplicity, we define $Z_t=(X_t,Y_t)$ as the pair of these two metrics at time $t$.
Let $\mathbb{Z}_{i}$ denote the sequence of metric records within each sliding window $i$. We can characterize the usage pattern of $\tau$ as a PDF on $\mathbb{Z}_{i}$, denoted as $P_{i}(Z)$, as illustrated in Figure \ref{fig:slid.1}. 

To accurately detect drifts, the pattern representation of microservices should remain robust during version upgrades under varying workloads, without being unaffected by the upgrade process. This entails two requirements. Firstly, the window size $W$ should be chosen to encompass the workload's  peaks and valleys, while ensuring that a single window predominantly contains the same version of the microservice. Secondly, the sliding step $S$ should span the entire duration of the upgrade process. The appropriate selection of $W$ and $S$ is discussed in Section \ref{sec:parameter_tuning}.

{
\begin{algorithm}[!htb]
\setlength{\abovedisplayskip}{3pt} 

\setlength{\belowdisplayskip}{3pt}
\caption{{$Permutation\_Test$}}
\label{alg:perm}
\KwIn{\textbf{$\mathbb{Z}_{i_{d}^{\prime}}~and~\mathbb{Z}_i$}: the reference data and the test data, \textbf{$\mu$}: the false positive rate to determine the confidence level, \textbf{$m$}: permutation test times;}
\KwOut{\textbf{$T_{\mu}$}: the threshold of $\hat{D}^2$ for hypothesis test;}

   \BlankLine
   $\mathbb{Z}_{all}=\mathbb{Z}_{i_{d}^{\prime}}\bigcup\mathbb{Z}_{i}$;
   
$\mathbb{T}_{cad}=[]$;~~~~~~\emph{// Candidate threshold values}

   \For{$i=1$ to $m$}{
$\mathbb{Z}^{\prime}_0,\mathbb{Z}^{\prime}_1 \leftarrow $ {randomly divide $\mathbb{Z}_{all}$ into 2 equal-size subsets;}

$\mathbb{T}_{cad}.add(\hat{D}^{2}(\mathbb{Z}^{\prime}_0,\mathbb{Z}^{\prime}_1))$; 
}
$T_{\mu}\leftarrow(1-\mu)\times 100$ percentile of $\mathbb{T}_{cad}$;
\end{algorithm}

}

Building upon this foundation, the detection of a pattern drift can be regarded as identifying the window $i_d$ in which the PDF exhibits a significant deviation compared to its previous state. Many existing studies, such as CPI2 \cite{zhang2013cpi2}, tend to predefine or profile the PDFs as specific types of distributions. Nevertheless, these schemes heavily rely on restrictive assumptions, thereby limiting their applicability in production environments with thousands of microservices. In contrast, we employ the least squares density difference ($LSDD$) \cite{LSDDCDT}, a widely recognized indicator that directly quantifies the difference in a machine-learning manner. For any two windows $u$ and $v$, the $LSDD$ is defined as follows:
$$
D^2(P_{u},P_{v})= \int \left(P_{u}(Z)-P_{v}(Z)\right)^2 d Z.
$$
As elaborated in LSDD-Inv \cite{LSDDINV}, we utilize Gaussian radial basis function (RBF) models to directly estimate $\hat{D}^2(\mathbb{Z}_u,\mathbb{Z}_v)$ between datasets $\mathbb{Z}_u$ and $\mathbb{Z}_v$.

\paragraph{Pattern Drift Detection Mechanism}
Based on the aforementioned formulations, we employ a hypothesis testing mechanism that relies on the $LSDD$ indicator to achieve online pattern drift detection.

For each microservice, let $i_{d}^{\prime}$ denote the time window when the most recent drift is detected. The detection process is executed at the conclusion of each window $i$. Each detection can be conceptualized as a hypothesis testing procedure with a confidence level $1-\mu$, where \textbf{$\mu$} represents the false positive rate to determine the confidence level. This testing can be formulated as follows:
$$
\begin{aligned}
H_0: P_{i^{\prime}_d}(Z) &= P_{i}(Z)\\
\end{aligned}
$$
Accepting $H_0$ implies that no drift has occurred in window $i$, while rejecting $H_0$ suggests the occurrence of a drift. As recommended in \cite{LSDDCDT}, we can accept or reject $H_0$ based on the following criterion:
$$\begin{cases}
    \text{Reject } H_0 &\text{If } \hat{D}^2(\mathbb{Z}_{{i}^{\prime}_d},\mathbb{Z}_i)>T_{\mu}\\
    \text{Accept } H_0 &\text{Otherwise}
    \end{cases}$$
where $T_{\mu}$ denotes the threshold of $\hat{D}^2$ in this test. At each detection, $T_{\mu}$ is determined using the $m$-times permutation test, as described in Algorithm \ref{alg:perm}.

{
\begin{algorithm}
\caption{Drift Detection Procedure}
\label{alg:total}
  \setlength{\abovedisplayskip}{3pt} 

  \KwOut{\textbf{$i_{d}$}: the window when the latest drift occurs;}
  \SetKwData{Left}{left}\SetKwData{This}{this}\SetKwData{Up}{up}
  \SetKwFunction{Union}{union}\SetKwFunction{FindCompress}{findcompress}
  \BlankLine                                    
    $counter\leftarrow 0$; \emph{// Counter for detected temporal drifts}
    
   $\mathbb{I}_{cad}\leftarrow[]$; \emph{// Candidates of identified window index}
   
\While{(true)}{
    {Obtain current window $i$, and datasets $\mathbb{Z}_{i_d^{\prime}}$ and $\mathbb{Z}_{i}$};
    
       $T_{\mu} \leftarrow Permutation\_Test(\mathbb{Z}_{i_d^{\prime}},\mathbb{Z}_{i})$;
       
          \If {$\hat{D}^{2}(\mathbb{Z}_{i_d^{\prime}},\mathbb{Z}_i)>T_{\mu}$}{
   $counter++,\;\mathbb{I}_{cad}.add(i)$;
   }
   \lElse{$counter \leftarrow 0, \mathbb{I}_{cad}\leftarrow []$}
   
   \If{$counter\geq \theta$}
    {$i_d=\min(\mathbb{I}_{cad})$;

   return $i_d$;} 
  } 
\end{algorithm}

}

Algorithm \ref{alg:total} presents the procedure for each detection cycle. At window $i$, a hypothesis test is constructed between $P_{i^{\prime}_d}$ and $P_i$ (lines $4$-$8$) to identify potential drifts. If drifts are detected in $\theta$ consecutive windows (lines $9$-$11$), it is inferred that a drift occurred. The earliest window $i_d$ where the hypothesis test is rejected is identified as the moment when the drift took place. All the metrics collected after window $i_d$ are utilized to accurately capture the latest usage pattern, which is employed for capacity adjustment, as elaborated in Section \ref{capacity_planner}.

\subsection{Capacity Adjuster}

The objective of capacity adjustment is to maintain the CPU utilization of microservices within the target upper limit $U^{*}$ under varying workloads, aiming to enhance performance stability and resource efficiency. To achieve this objective, the capacity adjuster needs to perform three tasks: workload forecasting, CPU usage pattern modeling, and resource capacity estimation. Following previous studies \cite{kae-informer,proscale}, capacity adjustment is conducted at regular intervals of $h_p$ hours.\footnote{In practical deployment, we set $h_p=1$, which effectively adapts to dynamic workloads while minimizing the overhead and service unavailability caused by frequent container initiation/termination.}


\subsubsection{Workload Forecasting}
Forecasting microservice workloads is a challenging task due to the intricate interplay of multiple components with diverse temporal dependencies \cite{kae-informer}. To ensure accurate predictions, we employ KAE-Informer \cite{kae-informer}, a state-of-the-art framework for workload forecasting. KAE-Informer disentangles the workload sequence into two distinct components that can be predicted: 1) the component representing the growth trend and dominant periodicity, and 2) the residual component encompassing long-range and local-context dependencies. To effectively model these dependencies, KAE-Informer proposes efficient architectures, exhibiting exceptional prediction capabilities when dealing with hour-level workload sequences of various microservices.

\subsubsection{Pattern Modeling}
As indicated by many studies \cite{luo2021characterizing,google_trace}, there exists a strong correlation between the CPU usage of microservices and the workload, with the former typically increasing as the workload rises. To capture this relationship with a balance of generality, accuracy and efficiency, we employ continuous regression functions. These functions can generally be approximated by a collection of polynomials based on Taylor's theorem \cite{sablik2000taylor}. Consequently, we represent the workload-usage relationship for each microservice as an ensemble of local linear models tailored to different workload ranges. Note that we refrain from training a single comprehensive model for all microservices, as frequent updates resulting from a large number of version upgrades can impede the stability of the model.

For microservice $\tau$, given its total workload $Y_t$ and normalized CPU usage $X_t$ at time $t$, we model the CPU usage pattern as $L$ segments of local linear models, denoted as $F$. These segments are determined by $L+1$ split points: $x_0<x_1<x_2<...<x_L$ as:
\begin{equation}   \label{eq:model}
        Y_t = F(X_t)=\sum_{l=1}^{L}\mathbf{sgn}(x_{l-1}\leq X_t <x_{l})\cdot(\alpha_{l} \cdot X_t + \beta_{l}),
\end{equation}
where $\mathbf{sgn}(\cdot)=1$ if true, otherwise 0.

We employ a statistical learning model known as generalized random forest (GRF) \cite{GRF}, which combines a collection of regression trees to estimate $F$. The selection of the optimal split points $\{x_{0},x_{1},...,x_{L}\}$ and the corresponding parameters $\{<\alpha_{l},\beta_{l}>|1\leq l \leq L\}$ is through an analysis of the leaf nodes within the regression trees\footnote{The complete learning procedure of GRF is elaborated at: { https://grf-labs.github.io/grf/articles/llf.html\#local-linear-forests-the-basics-1}}. 

\subsubsection{Resource Capacity Estimation}
\label{capacity_planner}

Estimating capacity involves two steps. Firstly, we estimate the CPU usage of each microservice in the upcoming $h_p$ hours based on the maximum predicted workload. For microservice $\tau$, when a new drift is detected at window $i_d$, we employ all the data collected after the drift to re-train the resource usage model $F$ as defined in Eq. (\ref{eq:model}). Let $X_{max}$ denote the maximum predicted workload of $\tau$, and we determine the maximum CPU usage as $Y_{max}=F(X_{max})$. Secondly, we estimate the resource requirements that can handle the maximum workload while holding utilization close to $U^{*}$ using the following equation:
\begin{equation}
    \label{eq:capacity_pred}
    {R}^{\prime} = \frac{Y_{max}}{U^{*}} \times (1+\psi),
\end{equation}
where $\psi$ denotes the margin to avoid surpassing $U^{*}$ and prevent resource over-utilization. Previous studies \cite{rzadca2020autopilot,deep-scaling} recommend $\psi$ to be within the range of $0.05\sim 0.15$.

\subsubsection{Adjustment Plan Generation}
Humas formulates an adjustment plan for each microservice $\tau$. 
Let ${r^{std}}$ represent the container CPU quota of $\tau$ deployed on the standard machines. Based on the most recent capacity $R^{\prime}$, the required number of standard containers ${n}^{\prime}$ for $\tau$ is determined as:
$$
{n}^{\prime} = \lceil\frac{R^{\prime}}{r^{std}}\rceil.
$$
The plan for container adjustment is calculated as follows:
\begin{equation}
\label{eq:delta}
\Delta{n} = {n}^{\prime} - {n}.
\end{equation}
where $n$ denotes the current number of containers. 

When deploying containers on machines of type $j$, the quota needs to be adjusted as follows:
\begin{equation}
\label{eq:heter_adjust}
r^{j} = RED^{j} \times {r^{std}}.
\end{equation}
where $RED^j$ is defined in Eq. (\ref{eq:red}). This quota adjustment takes into account the variation in work efficiency among heterogeneous machines to ensure system performance stability.

\section{Evaluation}
\subsection{Setup}
Following prior studies \cite{deep-scaling,Meta,Erms2023}, simulations were conducted by employing auto-scaling frameworks to dynamically adjust the resource capacity of the 50 microservices discussed in Section \ref{sec:moti}. The trace of these microservices was collected from Aug. 1 to Oct. 24, 2022, with a sampling interval of 1 minute. Throughout this period, a total of 922 version upgrades occurred. The simulated capacity adjustment began on Aug. 15. The scheduler utilizes the dominant resource fairness (DRF) policy \cite{drf} to allocate the containers of microservices across 1,500 $826X$ machines and 500 $816X$ machines, as described in Section \ref{sec:char_perf_gap}. In addition, the estimation of $LSDD$ values and the training of the capacity adjuster are executed on NVIDIA V100 GPU cards. 

In the simulation, each auto-scaling framework determines the adjustment plan $\Delta n$ of each microservice $\tau$ based on its own principle. At any given timestamp $t$, we compute the average container CPU usage of $\tau$ on each type of machine in the actual trace as the ground-truth container usage. When increasing or decreasing the $\Delta n$ containers of $\tau$, the scheduler disperses or reclaims them from $\Delta n$ machines with the lowest or highest CPU utilization respectively. This results in obtaining the adjusted capacity of $\tau$ on each type of machine. Thus, with the ground-truth container usage and the adjusted capacity, the corresponding CPU utilization can be estimated at each time $t$ using Eq. (\ref{eq:util_usage}).

Furthermore, for each microservice, we leverage the tail latency, i.e., the $95^{th}$ percentile of container latency, to evaluate the QoS satisfaction performance of auto-scaling frameworks. Inspired by the QoS-oriented auto-scaler Erms \cite{Erms2023}, we accomplish the profiling of tail latency distribution under varying RPS and resource utilization. Therefore, we yield the simulation of tail latency at each timestamp $t$ when executing each auto-scaling framework.

\begin{table}[tb]
\centering
\caption{ The parameter settings in evaluations}
\label{tab:parameter}
\vspace{-2mm}
{
\resizebox{0.48\textwidth}{!}{
\begin{tabular}{|c|c|c|}
 \hline
   {{Parameter}}&{Value}&{Description}
  \\
  \hline
   \multirow{6} *{} 
   $U^{*}$ &$40\% \sim 70\%$&The default target upper limit\\
    \cline{1-3}
$W$&$48$ hours& Size of the sliding window\\
    \cline{1-3}
    $S$&$8$ hours&Step of the sliding window\\
    \cline{1-3}
    $m$&$100$ &The sampling times in each permutation test\\
    \cline{1-3}
    $\mu$&$0.05$ &The false positive rate in hypothesis test\\
    \cline{1-3}
    $\theta$&$3$ &The threshold for tolerating temporal drift\\
    \cline{1-3}
    $h_{p}$&$1$ hour &Time interval of making auto-scaling decision\\
   \cline{1-3} 
    $\psi$& $0.08$&Safety margin for capacity estimation in Eq. (\ref{eq:capacity_pred})\\
    \cline{1-3} 
   \hline
\end{tabular}
}
\vspace{-4mm}
}
\end{table}

\begin{figure}[t]
     \centering

	\subfloat[Determining $U^{*}$ for $MS_1$]{\label{fig:u_limit.a}
		\begin{minipage}[t]{0.45\linewidth}
		\setlength{\abovecaptionskip}{-0.02cm}
		\centering
    \includegraphics[scale=0.171]{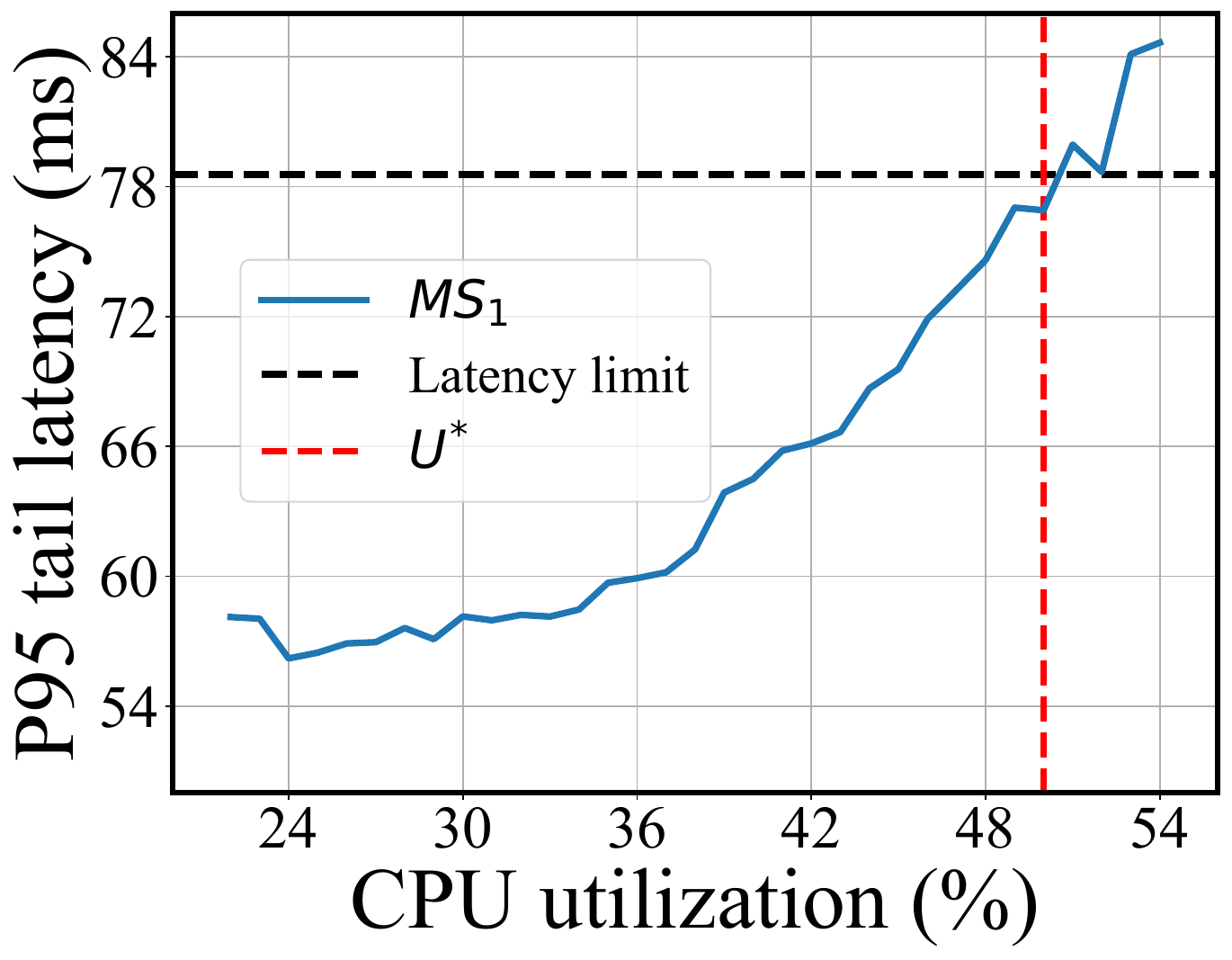}
		\end{minipage}
		}
  \subfloat[CDF of $U^{*}$ for $50$ microservices]{\label{fig:u_limit.b}
		\begin{minipage}[t]{0.45\linewidth}
		\setlength{\abovecaptionskip}{-0.02cm}
		\centering
    \includegraphics[scale=0.17]{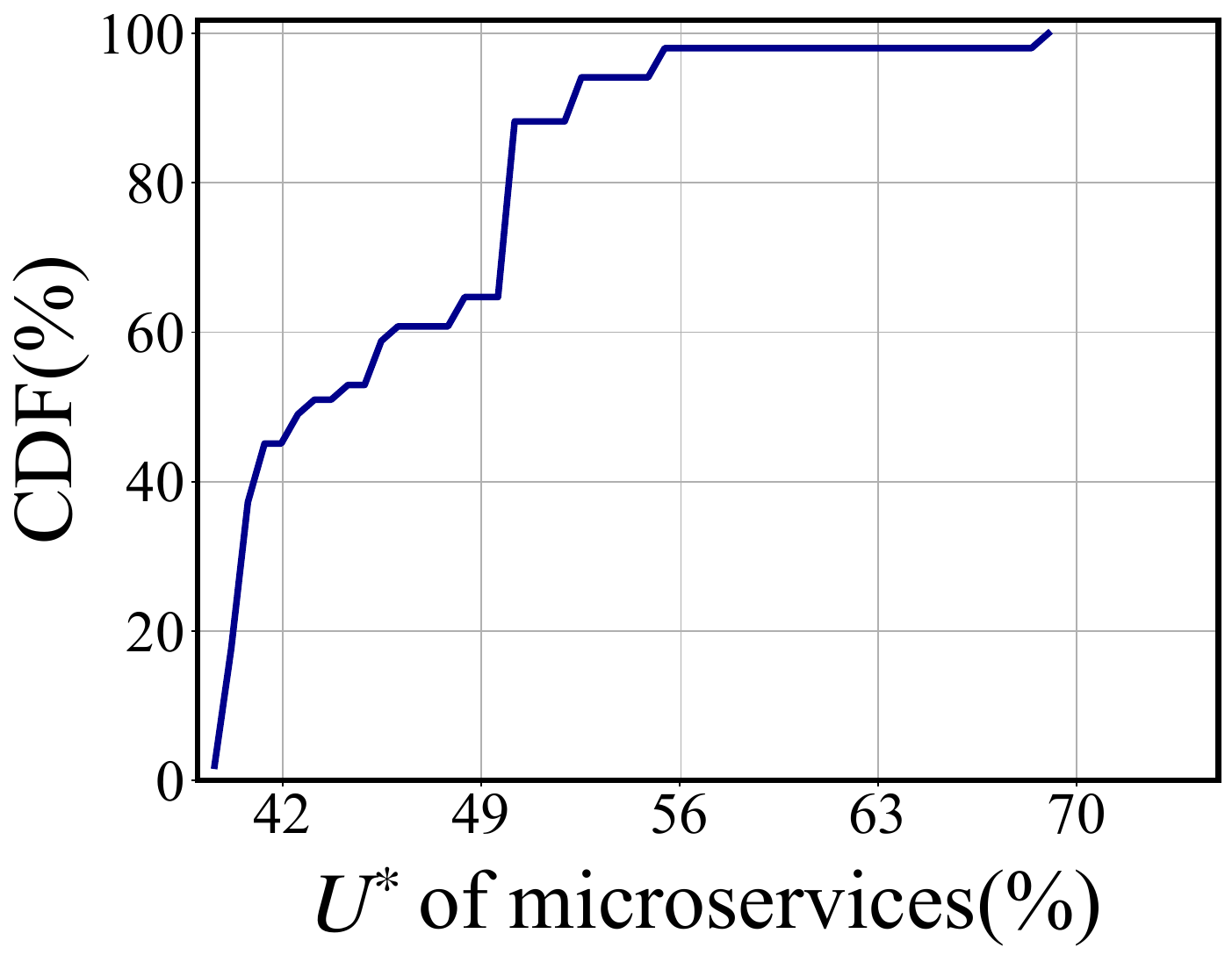}
		\end{minipage}
		}

    \caption{The parameter configuration of $U^{*}$ for $50$ large-scale microservices based on the QoS guarantee}
    \label{fig:overall_param_res}
 	\vspace{-0.5cm}
\end{figure}

\subsubsection{Parameter Settings}
\label{sec:parameter_setting}
We employ GRF \cite{GRF} to establish the parameters for capturing CPU usage patterns, and KAE-Informer\cite{kae-informer} to forecast workloads, as recommended in their respective papers. 
The configurations of the remaining hyperparameters for Humas are listed in Table \ref{tab:parameter}. In our experiments, for the detection procedure, we determine $m=100$ and $\mu=0.05$ based on research on the $LSDD$ indicator \cite{LSDDINV,LSDDCDT}. Additionally, we conduct experiments on the settings of $W$, $S$ and $\theta$, as discussed in Section \ref{sec:parameter_tuning}, and select the configuration that yielded the best performance.

Furthermore, in line with previous studies \cite{deep-scaling,sinan,Meta}, we set $U^{*}$ as the reference for ensuring the Qos guarantee. Specifically, for each microservice, we define the upper limit of tail latency ${\rho}^{*}$ as the ground-truth $99^{th}$ percentile obtained in the collected traces. Inspired by Meta \cite{Meta}, the corresponding $U^{*}$ is quantified based on the peak workload intensity, representing the maximum CPU utilization that keeps the $95^{th}$ percentile of tail latency below ${\rho}^{*}$. For example, Figure \ref{fig:u_limit.a} illustrates the profiling of $U^{*}=50$ for $MS_1$ as discussed. In addition, Figure \ref{fig:u_limit.b} presents the CDF of $U^{*}$ for the $50$ microservices, ranging from 40\% to 70\%.

\subsubsection{Metrics} 
We employ four metrics to assess the performance of auto-scaling frameworks, which can be categorized into two groups. Firstly, we adopt resource slack $(slack(\%))$ and average CPU allocation to evaluate resource efficiency. Secondly, we utilize two metrics to assess performance stability: the standard deviation (Std) of CPU utilization and the QoS violation rate (${Vio}(\%)$). The definitions of resource slack and violation rate are as follows.

\paragraph{$slack(\%)$}
Given a microservice $\tau$, let $\overline{U}_{\tau}$ and $U^{*}_{\tau}$ represent the average CPU utilization and the upper limit of $\tau$, respectively. In accordance with prior studies \cite{autoscaleopt,rzadca2020autopilot}, we define the resource slack of $\tau$ as: 
$$
 slack_{\tau}=\left(1-\frac{\overline{U}_{\tau}}{U^{*}_{\tau}}\right)\times 100\%
$$
According to \cite{autoscaleopt}, if $slack\geq0$, a lower value of $slack$ indicates that less resource capacity is wasted when processing the given workload, thereby leading to higher resource efficiency. Conversely, if $slack<0$, it signifies a performance violation. 


\paragraph{${Vio}(\%)$} 
At time $t$, let ${\rho_t}$ denote the tail latency of the containers belonging to $\tau$ having the upper limit of tail latency $\rho^{*}$. Following prior studies \cite{sinan,deep-scaling,Erms2023}, if $\rho_t>\rho^{*}$, we define the time point $t$ as a QoS violation occurrence for $\tau$. The violation rate of $\tau$ is computed as the ratio of the number of violation occurrences to the total number of time points.

\begin{figure*}[t]
     \centering
      \includegraphics[scale=0.23]{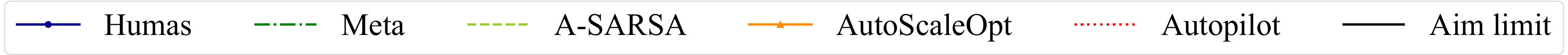}

    \vspace{-3mm} 
     \subfloat[$MS_0$]{\label{fig:trace.a}
		\centering \begin{minipage}[t]{0.2125\linewidth}
		\setlength{\abovecaptionskip}{-0.02cm}
		 \centering
		\includegraphics[scale=0.17] {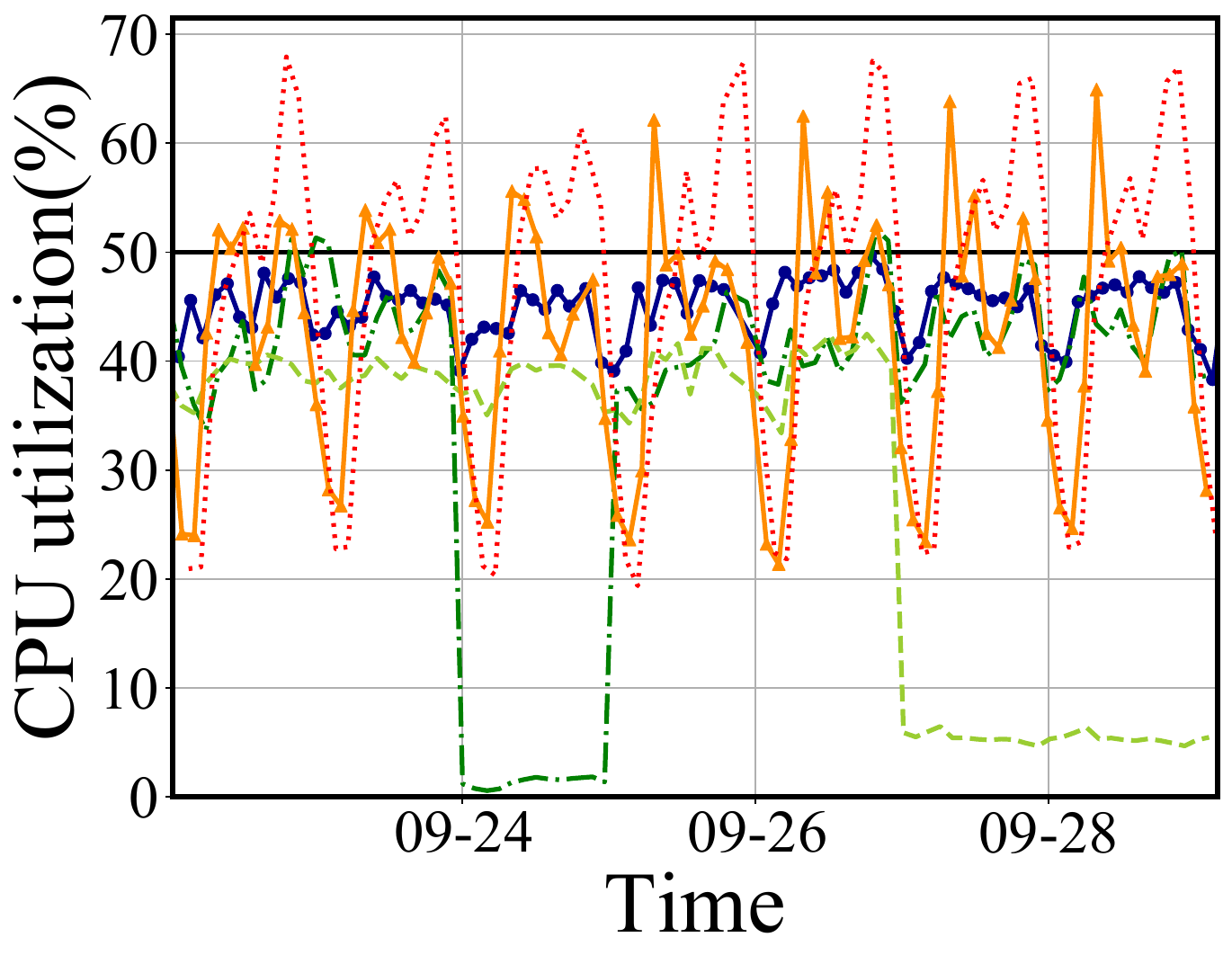}
		\end{minipage}
		}
  \hspace{0.01in}
	\subfloat[$MS_1$]{\label{fig:trace.b}
		\centering \begin{minipage}[t]{0.2125\linewidth}
		\setlength{\abovecaptionskip}{-0.02cm}
		 \centering
		\includegraphics[scale=0.17] {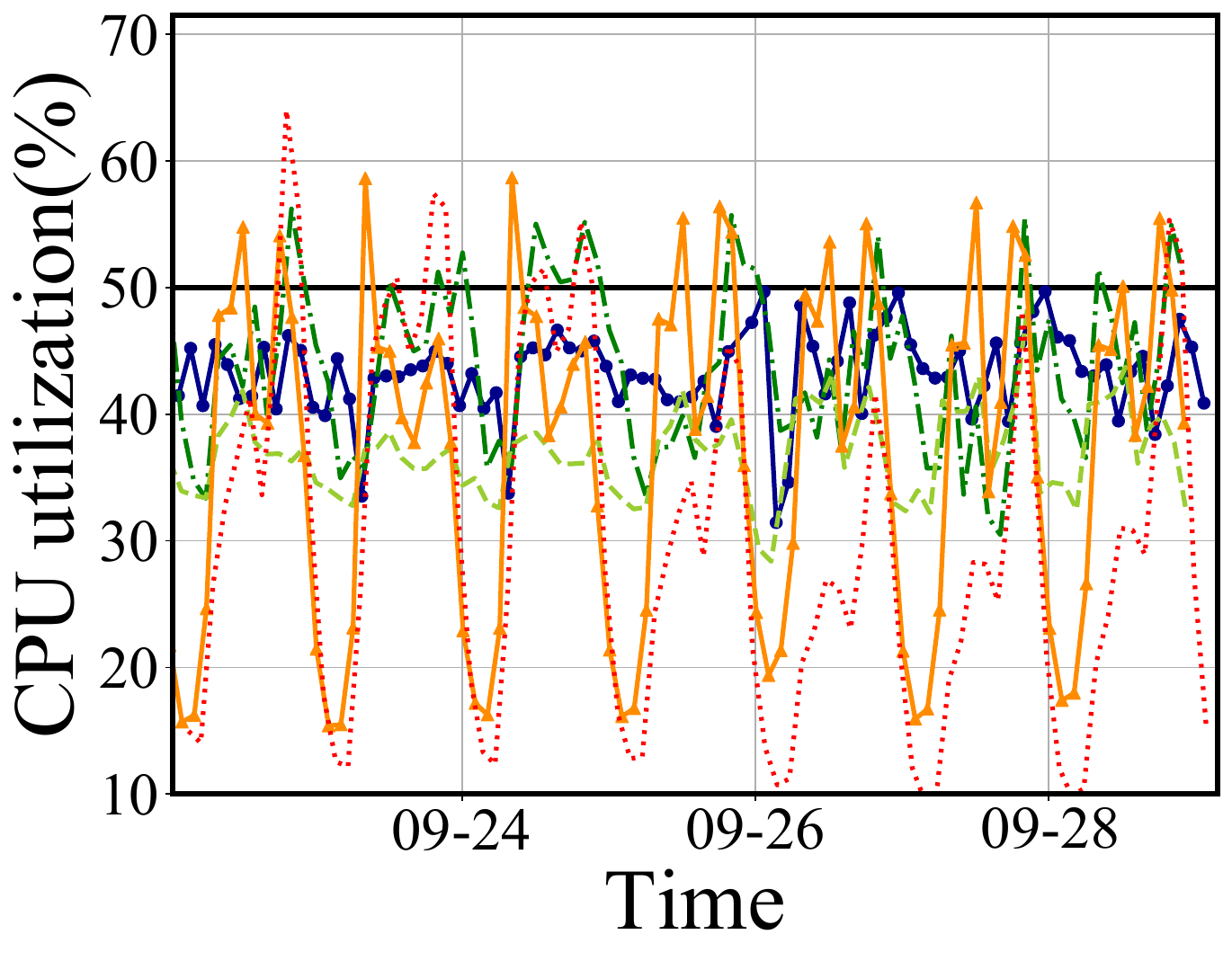}
		\end{minipage}
		}
\hspace{0.01in}
	\subfloat[$MS_2$]{\label{fig:trace.c}
		\begin{minipage}[t]{0.2125\linewidth}
		\setlength{\abovecaptionskip}{-0.02cm}
		\centering
		\includegraphics[scale=0.17]{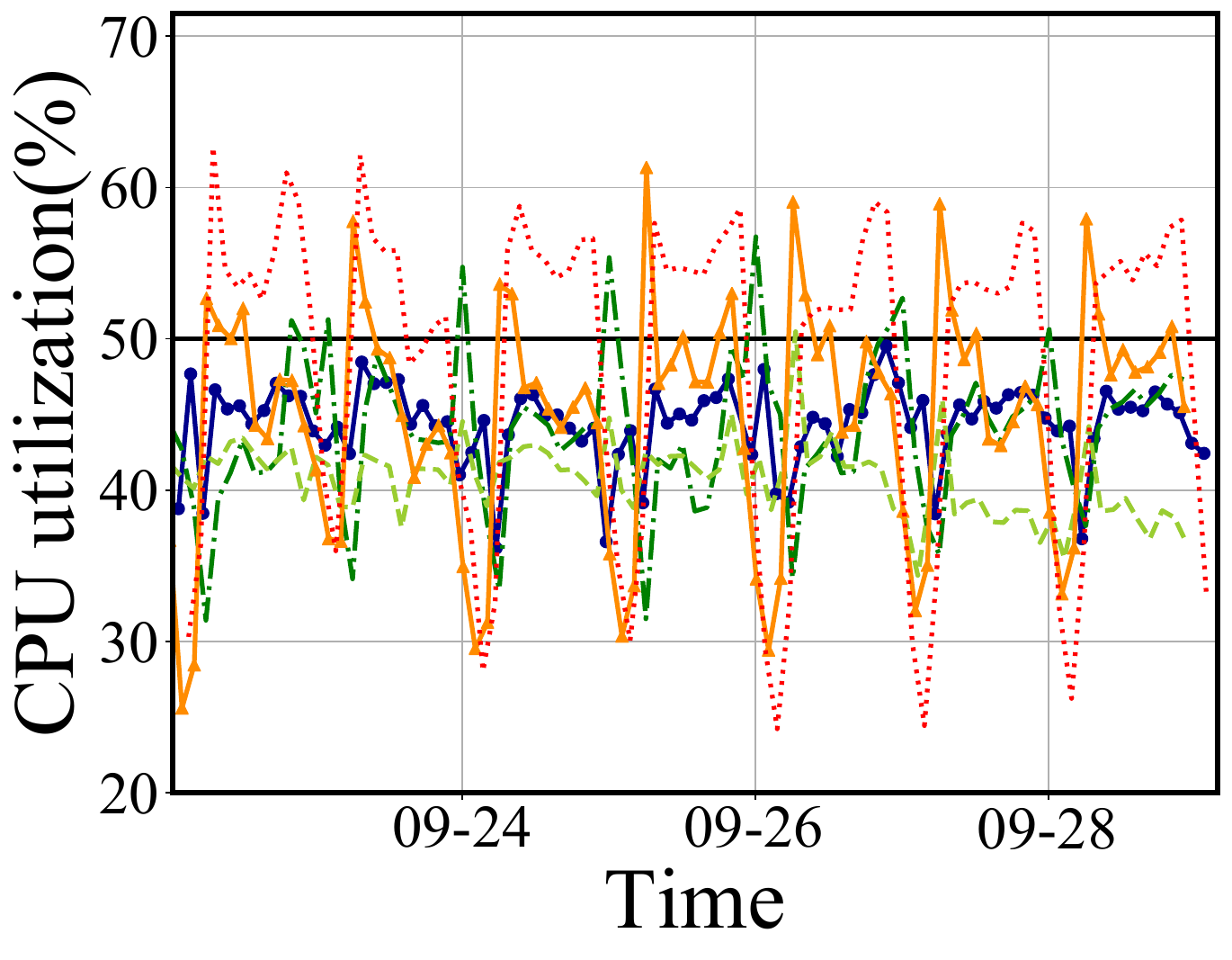}
		\end{minipage}
		}
  \hspace{0.01in}
  \subfloat[$MS_3$]{\label{fig:trace.d}
		\begin{minipage}[t]{0.2125\linewidth}
		\setlength{\abovecaptionskip}{-0.02cm}
		\centering
		\includegraphics[scale=0.17]{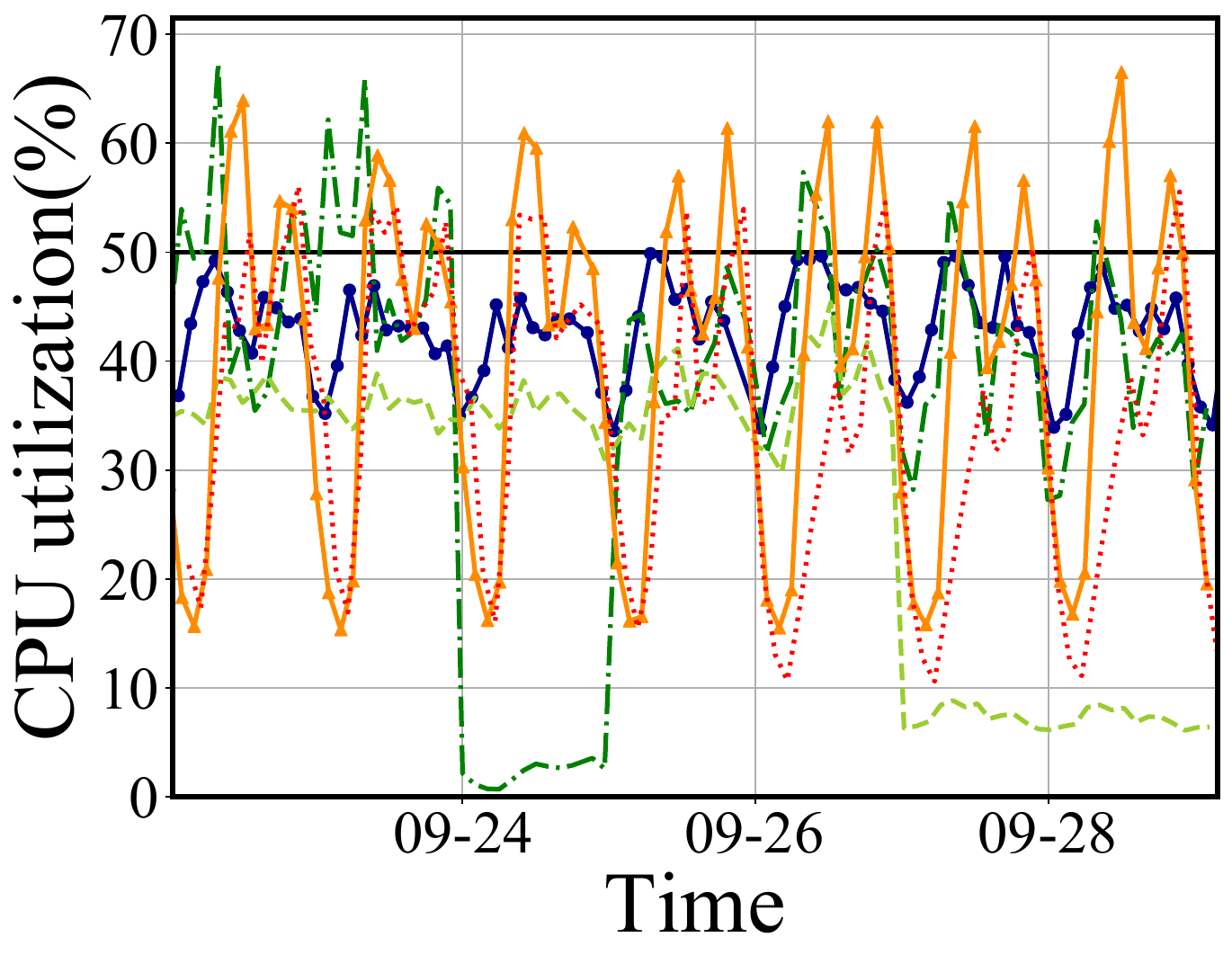}
		\end{minipage}
		}
	\vspace{-2mm}
    \caption{7-day CPU utilization traces of the four representative microservices when executing the five auto-scaling approaches}
    \label{fig:trace}
 	\vspace{-0.4cm}
\end{figure*}

\subsubsection{Baselines} 
We compare Humas with four state-of-the-art baselines. 1) Autopilot \cite{rzadca2020autopilot} is a statistic-based approach proposed by Google. It enhances resource efficiency by utilizing the weighted histogram of CPU usage in prior time windows to estimate the container CPU quota. 2) AutoScaleOpt \cite{autoscaleopt} is a rule-based approach employed in Amazon. It increases CPU capacity by $10\%$ and $30\%$ when utilization falls within the range of $[50\%, 70\%)$ and $[70\%, 100\%]$, respectively. Conversely, it reduces CPU capacity by $10\%$ and $30\%$ when utilization is in the range of $[30\%, 40\%)$ and $[0\%, 30\%)$, respectively. 3) We consider two proactive approaches: a) A-SARSA \cite{ASARSA} employs ARIMA and DNN to predict workload and learn CPU usage patterns, respectively. It also utilizes reinforcement learning (RL) to estimate resource capacity to avoid violations; and b) Meta \cite{Meta} uses LSTM, ANP and RL to forecast workload, learn usage patterns, and estimate CPU capacity respectively, aiming to enhance performance stability.

\subsection{Overall Performance}
Table \ref{tab:overall} presents the performance evaluation of the 50 microservices across the execution of the five frameworks.

\begin{table}[tb]
\centering
\caption{ The performance of the five frameworks and raw workload traces on the 50 large-scale microservices}
\label{tab:overall}
\vspace{-2mm}
\resizebox{0.498\textwidth}{!}{
\begin{threeparttable}
\begin{tabular}
{|c|c|c|c|c|c|c|c|c|c|c|}
 \hline
   {\multirow{3}*{Frameworks}}&\multicolumn{2}{c|}{Resource Efficiency}&\multicolumn{4}{c|}{Performance Stability}
  \\
  \cline{2-7}
  &\multirow{2} *{$Slack(\%)$}
    &\multirow{2} *{Capacity (Cores)} &\multicolumn{3}{c|}{Utilization Std}&\multirow{2} *{$Vio(\%)$}\\
    \cline{4-6}
    &&&{Total}&{$826X$}&{$816X$}&\\
    \cline{2-7}
  \hline
   \multirow{6} *{}
   Autopilot
   &$16.63$&$23,832.6$&\textcolor{black}{${14.37}$} 
   &\textcolor{black}{${15.31}$}
   &\textcolor{black}{$18.42$} 
   &{$9.14$}  
   \\
   \cline{1-7}
   AutoScaleOpt
   &{$17.05$}
   &$25,356.8$&\textcolor{black}{$12.62$}
   &\textcolor{black}{$16.09$}
   &\textcolor{black}{$20.19$}
   &$6.31$
   \\
    \cline{1-7}
   A-SARSA
   &{$25.46$}
   &$32,090.3$&\textcolor{black}{$8.91$}
   &\textcolor{black}{$12.39$}&{$18.90$}
   &$2.67$\\
   \cline{1-7} 
   Meta
   &$17.67$
   &$30,245.8$&\textcolor{black}{$7.08$}
   &\textcolor{black}{$11.49$}
   &\textcolor{black}{$16.07$}
   &{$2.79$}\\
   \cline{1-7}
   Raw Traces
   &{$52.87$}
   &$44,283.4$&\textcolor{black}{$8.16$}
   &\textcolor{black}{$7.96$}
   &\textcolor{black}{$10.62$}
   &$\mathbf{0.85}$
   \\
   \cline{1-7}
   Humas(ours)
   &{$\mathbf{11.58}$}
   &$\mathbf{23,580.2}$&\textcolor{black}{$\mathbf{3.68}$}
   &\textcolor{black}{$\mathbf{3.63}$}
   &\textcolor{black}{$\mathbf{4.32}$}
    &$1.28$
  \\
   \hline
\end{tabular}
\begin{tablenotes}
  \small
  \item [1] The value of $slack$, utilization Std and $Vio(\%)$ are the average of metric values on the 50 microservices weighted by their average CPU capacity respectively.  
\end{tablenotes}
\end{threeparttable}
}
\vspace{-4mm}
\end{table}

\begin{figure*}[t]
     \centering
\subfloat[Average  $slack$]{\label{fig:overall.c}
		\begin{minipage}[t]{0.225\linewidth}
		\setlength{\abovecaptionskip}{-0.02cm}
		\centering
    \includegraphics[scale=0.17]{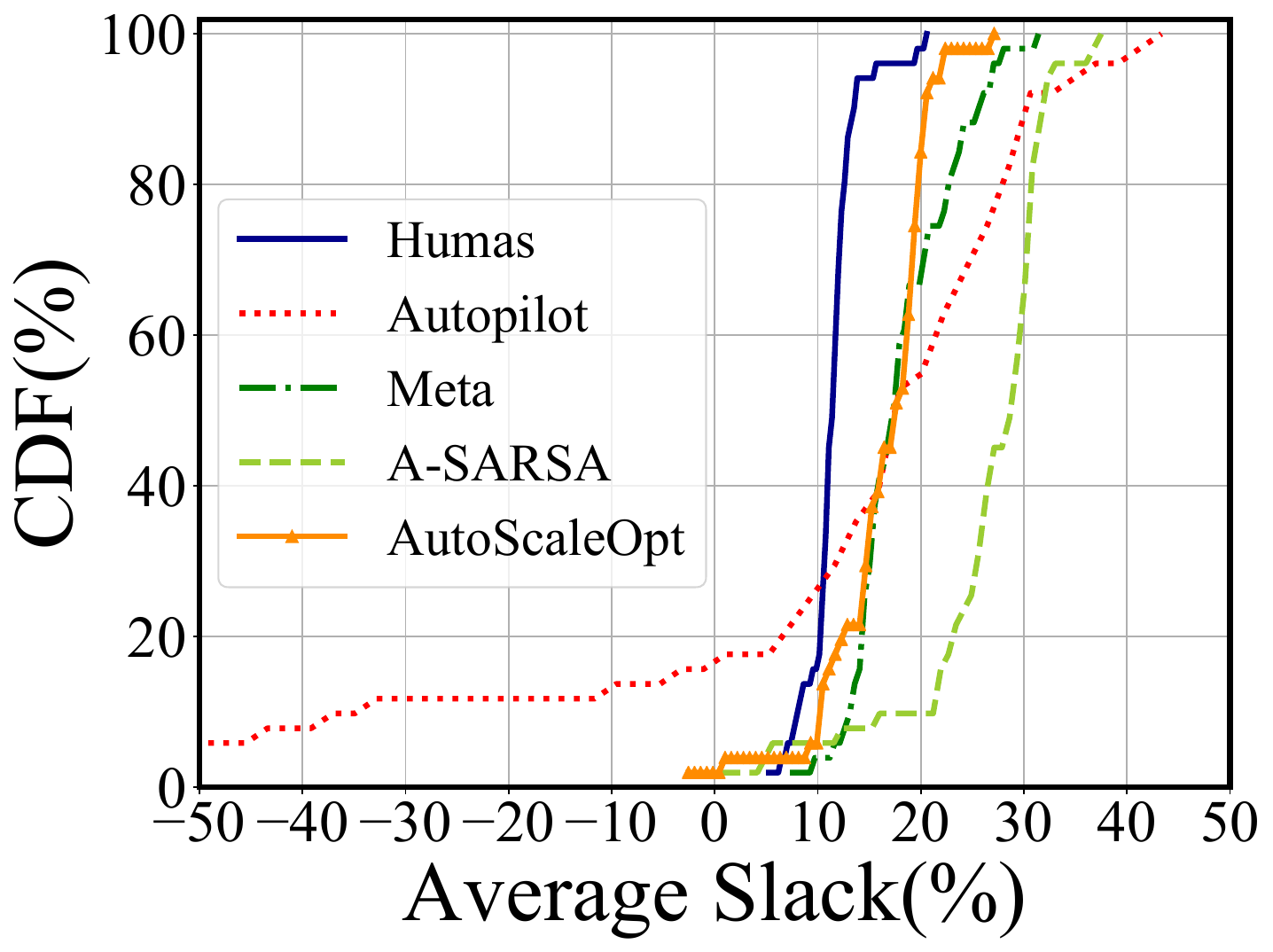}
		\end{minipage}
		}
  \hspace{0.01in}
	\subfloat[CPU utilization Std]{\label{fig:overall.b}
		\begin{minipage}[t]{0.225\linewidth}
		\setlength{\abovecaptionskip}{-0.02cm}
		\centering
    \includegraphics[scale=0.17]{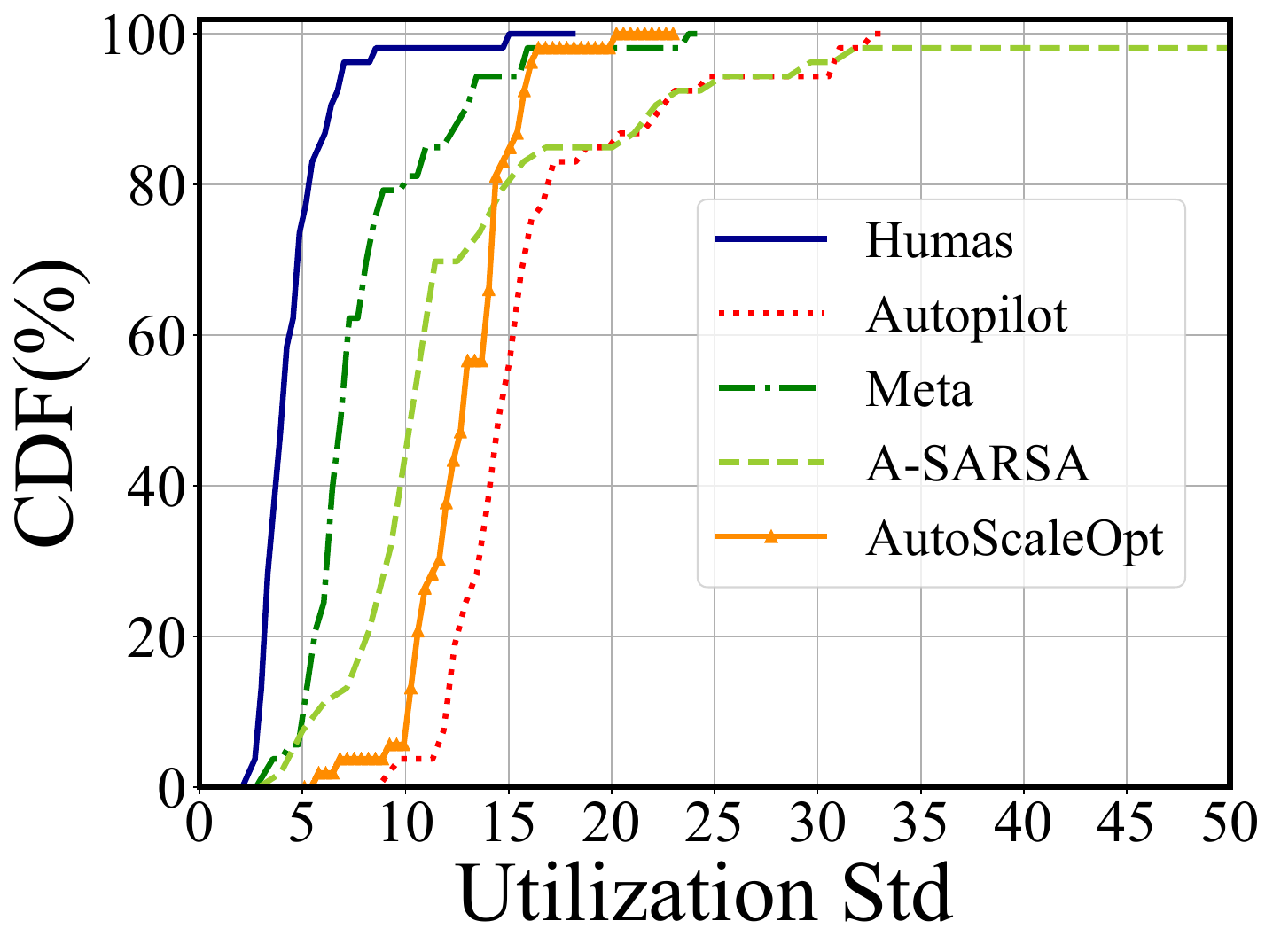}
		\end{minipage}
		}
  \hspace{0.02in}
  \subfloat[Performance violation rate]{\label{fig:overall.d}
		\begin{minipage}[t]{0.225\linewidth}
		\setlength{\abovecaptionskip}{-0.02cm}
		\centering
    \includegraphics[scale=0.17]{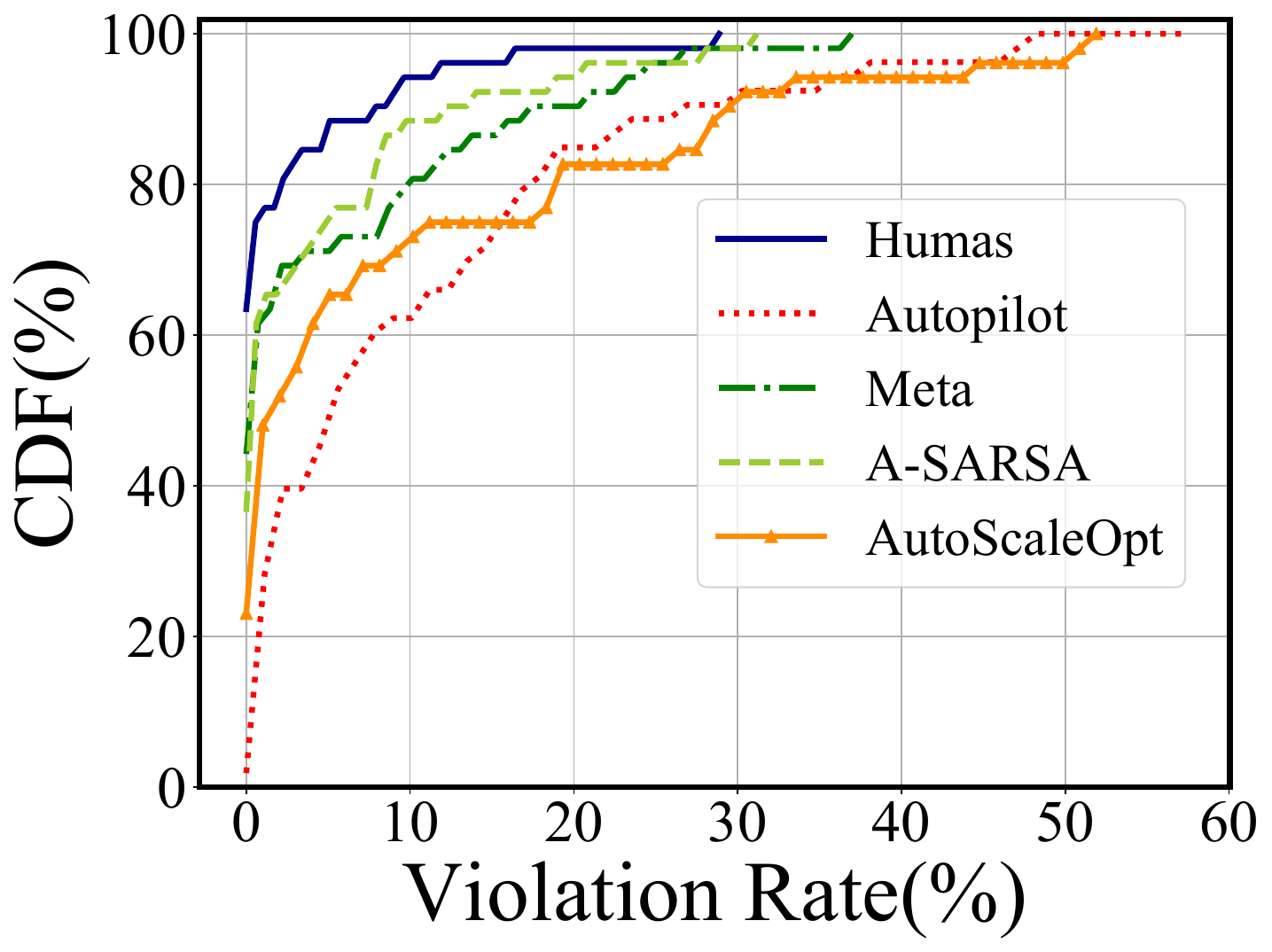}
		\end{minipage}
		}
  \hspace{0.01in}
  \subfloat[CPU usage change rate of upgrades]{\label{fig:overall.e}
		\begin{minipage}[t]{0.245\linewidth}
		\setlength{\abovecaptionskip}{-0.02cm}
		\centering
 \includegraphics[scale=0.17]{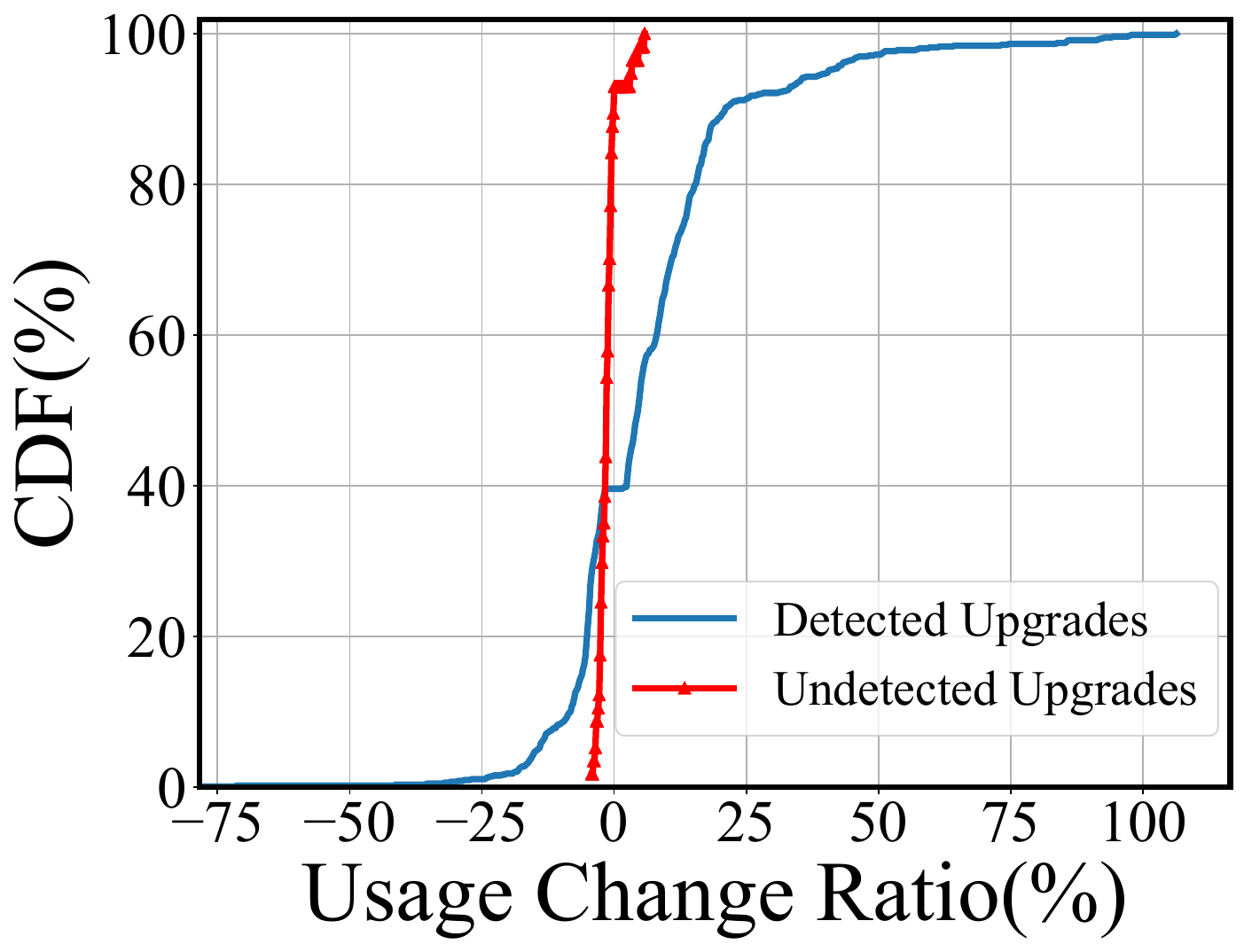}
		\end{minipage}
		}
	\vspace{-0.2cm}
    \caption{The CDF results of the performance on $50$ large-scale microservices}
    \label{fig:overall_res}
 	\vspace{-0.4cm}
\end{figure*}

\paragraph{Resource Efficiency} First, compared to Autopilot, AutoScaleOpt, A-SARA and Meta, Humas demonstrates an average reduction in resource slack by about 30.4\%, 32.1\%, 54.5\% and 34.5\%, respectively. For example, Figure \ref{fig:trace} shows that Humas consistently maintains the utilization of $MS_0$ $\sim$ $MS_3$. Moreover, considering all 50 microservices, Figure \ref{fig:overall.c} indicates that 
Humas reduces the $95^{th}$-percentile ($P95$) value of resource slack by about 60.2\%, 33.3\%, 57.2\% and 48.5\%, when compared to the four baselines, respectively. 

On one hand, Humas effectively eliminates instances of severe resource under-provisioning ($slack<0$) that are observed in Autopilot. On the other hand, Humas successfully addresses significant resource waste ($slack>50\%$), thereby reducing the average CPU capacity allocation by about 46.8\% compared to the raw traces. In summary, Humas can effectively enhance the resource efficiency of microservices by mitigating under- and over-provision issues.

\paragraph{Performance Stability} 
 As presented in Table \ref{tab:overall}, Humas demonstrates an average reduction in the Std of CPU utilization by about 74.4\%, 70.8\%, 58.7\%, and 48.0\% compared with Autopilot, AutoScaleOpt, A-SARA, and Meta, respectively. In addition, Humas exhibits a reduction in the overall violation rate by about 86.0\%, 79.7\%, 52.1\%, and 54.1\%, respectively. Moreover, Figure \ref{fig:trace} visually shows the superior stability of Humas across the four large-scale microservices, significantly mitigating performance fluctuations.

We also analyze all 50 microservices. 
Firstly, in comparison to the best baseline Meta, Humas reduces the $P50$ and $P95$ utilization Std by about 43.4\% and 56.6\% respectively, as depicted in Figure \ref{fig:overall.b}. Secondly, Figure \ref{fig:overall.d} indicates that, compared with the best baseline A-SARA, Humas reduces the $P95$ violation rate by about 34.8\%. Thirdly, as presented in Table \ref{tab:overall}, compared with Autopilot and Meta, Humas demonstrates an average reduction of around 5.4\% and 43.9\% respectively in the utilization Std difference between containers deployed on $826X$ and $816X$ machines. 
These outcomes validate the effectiveness of Humas in enhancing performance stability, particularly in the presence of hardware heterogeneity. 


The performance improvements of Humas can be attributed to three factors. 
Firstly, Humas achieves the consistent and precise learning of the latest CPU usage patterns by promptly detecting drifts. Secondly, the resource normalization based on $RED$ eliminates the influence of hardware heterogeneity on performance. Thirdly, KAE-Informer enhances the long-term workload forecasting capability. These three factors collectively enable Humas to accurately estimate resource capacity.


\subsection{Ablation Studies}
\label{sec:ablation}

We propose three variants of Humas: 1) $\text{Hmas}_{D8}$ and 2) $\text{Hmas}_{D2}$, which exclude the drift detector from Humas and update performance patterns every 8 days and 2 days, respectively; and 3) Umas, which eliminates the resource normalizer from Humas. The fixed update intervals of 2 and 8 days are determined based on the average and $P95$ of upgrade frequency. The overall results of the ablation studies are presented in Table \ref{tab:abs_overall}.

\begin{table}[tb]
\centering
\caption{ The overall performance in ablation studies}
\label{tab:abs_overall}
\vspace{-2mm}
{
\resizebox{0.485\textwidth}{!}{
\begin{tabular}{|c|c|c|c|c|c|c|}
 \hline
   {\multirow{3}*{Variants}}&\multicolumn{2}{c|}{Resource Efficiency}&\multicolumn{4}{c|}{Utilization Stability}
  \\
  \cline{2-7}
    &\multirow{2}*{$Slack(\%)$}&\multirow{2}*{Capacity(Cores)}&\multicolumn{3}{c|}{Utilization Std}&\multirow{2}*{${Vio}(\%)$}\\
    \cline{4-6}
    &&&{Total}&{$826X$}&{$816X$}&\\
    \cline{2-7}
  \hline
   \multirow{4} *{}
   $\text{Hmas}_{D8}$&$13.28$&${24,035.5}$&\textcolor{black}{${4.49}$}&\textcolor{black}{${4.42}$}&\textcolor{black}{${5.19}$}&{$1.74$}\\
   \cline{1-7}
   $\text{Hmas}_{D2}$&{$13.57$}&{$24,111.5$}&\textcolor{black}{$4.05$}&\textcolor{black}{$3.99$}&{$4.70$}&$1.66$\\
   \cline{1-7} 
    Umas&{$13.31$}&{$23,971.4$}&\textcolor{black}{$3.95$}&{$10.01$}&{$14.09$}&$2.48$\\
   \cline{1-7} 
    Humas(ours)
   &{$\mathbf{11.58}$}
   &$\mathbf{23,580.2}$&\textcolor{black}{$\mathbf{3.68}$}
   &\textcolor{black}{$\mathbf{3.63}$}
   &\textcolor{black}{$\mathbf{4.32}$}
    &$\mathbf{1.28}$\\
   \hline
\end{tabular}
}
}
\vspace{-4mm}
\end{table}

\begin{table}[bt]
\centering
\caption{ The overall performance of workload forecasters}
\label{tab:through_forecast}
\vspace{-2mm}
{
\resizebox{0.35\textwidth}{!}{
\begin{tabular}{|c|c|c|}
 \hline
   {\multirow{1}*{Models}}

&\multicolumn{1}{|c|}{RMSE}&\multicolumn{1}{|c|}{MAE}\\
\cline{2-3} 
  \hline
   \multirow{6} *{}
    KAE-Informer in Humas (ours)&$\mathbf{0.0239}$&$\mathbf{0.0148}$\\ 
    \cline{1-3}
   Arima in A-SARSA&$0.0533$&$0.0368$\\
   \cline{1-3}
   LSTM with attention layers in Meta&${0.1545}$&$0.1045$\\
   \cline{1-3}
Prophet \cite{prophet}&${0.0651}$&$0.0492$\\
   \cline{1-3} 
   Informer\cite{zhou2021informer}&$0.0350$&$0.0213$\\
   \cline{1-3}
   \hline
\end{tabular}}
\vspace{-0.3cm}
}
\end{table}

\subsubsection{Workload Forecaster}
A comparison of four baseline forecasters, as outlined in Table \ref{tab:through_forecast}, shows that Humas decreases the normalized root mean square error (RMSE) and mean absolute error (MAE) of workload prediction by about 31.7\% and 30.5\% respectively. This improvement can be attributed to the effective disentanglement of patterns and the extraction of dependencies accomplished by KAE-Informer. With accurate workload forecasting, Humas can further optimize resource capacity estimation.

\begin{table}[tb]
\centering
\caption{{Statistics of three e-commerce microservices}}
\label{tab:key_ms2}
\vspace{-2mm}
{
\resizebox{0.48\textwidth}{!}{
\begin{tabular}{|c|cc|c|c|c|c|c|}
 \hline
  {\multirow{2}*{{Name}}}&\multicolumn{2}{c}{\multirow{2}*{{Functionality}}}&\multicolumn{5}{|c|}{{CPU Capacity (Cores)}}\\
  \cline{4-8}
  &&&{{Total}}&{{$826X$}}&{{$816X$}}&{{$836X$}}&{{$ARMv9$}}\\
  \cline{1-8}
   \multirow{3} *{}
   \multirow{1}*{{$MS_4$}} &\multicolumn{2}{c|}{{Order confirmation}}&\multirow{1}*{{$43,672$}}&\multirow{1}*{{$12,880$}}&\multirow{1}*{{$2,200$}}&\multirow{1}*{{$12,144$}}&\multirow{1}*{{$16,448$}}\\
   \cline{1-8} 
    {$MS_5$}&  \multicolumn{2}{c|}{{Cart management}} &{$48,192$}&{$9,968$}&{$1,072$}&{$16,192$}&{$20,960$}\\
    
    \cline{1-8}
    {$MS_6$}&  \multicolumn{2}{c|}{{Discount management}}&{$43,344$}&{$9,328$}&{$1,680$}&{$15,920$}&{$16,416$}\\
    \cline{1-8}
\cline{1-8}
   \hline
\end{tabular}
}
\vspace{-0.2cm}
}
\end{table}

\subsubsection{Heterogeneity-aware Resource Normalizer}

\label{sec:heter_effect_eval}



Table \ref{tab:abs_overall} illustrates that, in comparison to Umas, Humas significantly reduces the Std of CPU utilization by about 63.7\% and 69.3\% on $826X$ and $816X$ machines respectively. This validates the effectiveness of the $RED$ measurement. 
Additionally, considering the fact that a majority of the 50 microservices exhibit lower work efficiency on $816X$ machines (Section \ref{sec:char_perf_gap}), the majority of violations in Umas occur on $816X$ machines, accounting for about 64.4\%. By adjusting the CPU quota of containers on $816X$ machines using Eq. (\ref{eq:heter_adjust}) to alleviate excessively high CPU utilization resulting from efficiency disparities, Humas reduces the overall violation rate by about 48.4\% compared with Umas.

{To further verify the adaptability of Humas in highly heterogeneous environments, we conduct an evaluation on three additional large-scale microservices that encompass the core functionalities of a leading e-commerce platform, as outlined in Table \ref{tab:key_ms2}. These microservices are deployed in a separate data center equipped with four distinct CPU models: 1) $826X$, 2) $816X$, 3) Intel Xeon Platinum $836X$ (Ice Lake model), and 4) customized $ARMv9$. Following the parameter configuration method presented in Section \ref{sec:parameter_setting}, we set the utilization limit $U^{*}$ within the range $40\%\sim55\%$, and with $826X$ as the standard type.}

{Table \ref{tab:more_heter} presents the results of Humas and Umas on the three e-commerce microservices. It can be observed that, in comparison to Umas, Humas significantly reduces the Std of CPU utilization by about 57.7\%, 55.5\%, 41.8\% and 58.4\% on $826X$, $816X$, $836X$ and $ARMv9$ machines, respectively. Remarkably, Humas consistently reduces the Std of utilization for each microservice across all four machine types, thus validating the effectiveness of the resource normalization mechanism of Humas. Furthermore, Humas achieves a substantial reduction of approximately 76.9\% in overall QoS violation compared to Umas. This demonstrates the effectiveness of the container CPU quota adjustment, as presented in Eq.(\ref{eq:heter_adjust}), in mitigating the impact of efficiency difference on service performance across the highly heterogeneous machine types. Collectively, these results indicate that Humas exhibits robust generalization capabilities in adapting to heterogeneous environments.}

\begin{table}[tb]
\centering
\caption{ {Performance on the four types of machines}}
\label{tab:more_heter}
\vspace{-2mm}
{
\resizebox{0.485\textwidth}{!}{
\begin{tabular}{|c|c|c|c|c|c|c|c|c|}
 \hline
   &{\multirow{3}*{{Variants}}}&\multicolumn{1}{c|}{{Resource Efficiency}}&\multicolumn{6}{c|}{{Utilization Stability}}
  \\
  \cline{3-9}
    &&\multirow{2}*{{$Slack(\%)$}}&\multicolumn{5}{c|}{{Utilization Std}}&\multirow{2}*{{${Vio}(\%)$}}\\
    \cline{4-8}
    &&&{{Total}}&{{$826X$}}&{{$816X$}}&{{$836X$}}&{{$ARMv9$}}&\\
    \cline{3-9}
  \hline
   {\multirow{2}*{{$MS_4$}}}&\multirow{2} *{}
    {Umas}&{{$16.72$}}&{{$3.92$}}&{{$6.51$}}&{{$6.84$}}&{$5.03$}&{$6.85$}&{$2.15$}\\
   \cline{2-9} 
    &{Humas}
   &{{$\mathbf{13.32}$}}&{$\mathbf{3.02}$}&{$\mathbf{3.17}$}
   &{$\mathbf{3.26}$}
   &{$\mathbf{3.52}$}
    &{$\mathbf{3.48}$}&{$\mathbf{0.77}$}\\
    \cline{1-9}
    {\multirow{2}*{{$MS_5$}}}&\multirow{2} *{}
    {Umas}&{{$13.60$}}&{{$\mathbf{4.32}$}}&{$10.65$}&{$11.18$}&{$8.08$}&{$10.98$}&{$2.86$}\\
   \cline{2-9} 
   & {Humas}
   &{$\mathbf{13.58}$}
   &{$4.52$}&{$\mathbf{4.52}$}
   &{$\mathbf{4.67}$}
   &{$\mathbf{4.74}$}
    &{$\mathbf{4.63}$}&{$\mathbf{0.71}$}\\
    \cline{1-9}{\multirow{2}*{{$MS_6$}}}&\multirow{2} *{}
    {Umas}&{$\mathbf{13.35}$}&{$3.30$}&{$7.99$}&{$6.75$}&{$6.53$}&{$8.93$}&{$3.01$}\\
   \cline{2-9} 
    &{Humas}
   &{${13.81}$}
   &{$\mathbf{2.96}$}&{$\mathbf{2.95}$}
   &{$\mathbf{3.13}$}
   &{$\mathbf{3.17}$}
    &{$\mathbf{3.03}$}&{$\mathbf{0.37}$}\\
    \cline{1-9} {\multirow{2}*{{Overall}}}&\multirow{2} *{}
   
    {Umas}&{$14.53$}&{$3.86$}&{$8.46$}&{$8.36$}&{$6.60$}&{$8.99$}&{$2.68$}\\
   \cline{2-9} 
   & {Humas}&{$\mathbf{13.57}$}
   &{$\mathbf{3.53}$}&{$\mathbf{3.58}$}
   &{$\mathbf{3.72}$}
   &{$\mathbf{3.84}$}
    &{$\mathbf{3.74}$}&{$\mathbf{0.62}$}\\
   \hline
\end{tabular}
}
}
\vspace{-2mm}
\end{table}

\subsubsection{Upgrade-aware Pattern Drift Detection}

\begin{table}[tb]

\centering
\caption{Overall performance of pattern drift detection}
\label{tab:drift_perform} 

{
\resizebox{0.48\textwidth}{!}{
\begin{threeparttable}
\begin{tabular}{|c|c|c|c|c|c|c|}
 \hline
   {\multirow{2}*{Approach}}&\multicolumn{3}{c|}{Overall Results }&\multicolumn{3}{c|}{Overall Performance}
  \\
  \cline{2-7}
    &\multicolumn{1}{c|}{TDD}&\multicolumn{1}{c|}{FDD}&\multicolumn{1}{c|}{DD}&\multicolumn{1}{c|}{Precision (\%)}&\multicolumn{1}{c|}{Recall (\%)}&\multicolumn{1}{c|}{Overhead (s)}\\
    \cline{1-7}
  \hline
   \multirow{1}*{$LSDD$}&$793$&{$77$}&$870$&{$94.40$}&{$91.04$}&{0.0146}\\
   \cline{1-7}
   \multirow{1}*{$MMD$}&$784$&$85$&$869$&$93.33$&$90.22$&{0.2160}\\
    \cline{1-7}
   \multirow{1}*{$CMMD$}&$803$&$72$&$875$&$95.60$&$91.77$&{31.5106}\\
   \hline
\end{tabular}
 \begin{tablenotes}
  \small
  \item [1] TDD, FDD and DD denote true detected drifts, false detected drifts, and detected drifts, respectively. 
  \item [2] Precision $=\frac{\text{TDD}}{\text{DD}}\times100\%$, Recall$=\frac{\text{TDD}}{\text{Upgrades}}\times100\%$.   
\end{tablenotes}
\end{threeparttable}
}
}
\vspace{-4mm}
\end{table}

\paragraph{Detection Indicator}
We compare the $LSDD$ indicator in Humas with two baselines called $MMD$ \cite{MMD-drift} and $CMMD$ \cite{pmlr-v162-cobb22a}. These baselines substitute LSDD in Humas with two commonly employed indicators, namely maximum mean discrepancy (MMD) \cite{MMD-drift} and context-aware MMD \cite{pmlr-v162-cobb22a}, respectively. 
For each drift detected at window $i_d$, we classify it as true if the corresponding upgrade occurs in window $i_d$ or $i_d-1$. Otherwise, we consider it as a false positive. Using the actual upgrades as the ground truth, we employ three metrics to assess the performance of drift detection for each microservice: precision, recall, and detection overhead.

Table \ref{tab:drift_perform} presents the overall performance evaluation of $LSDD$, $MMD$, and $CMMD$ in terms of drift detection on the dataset comprising the $50$ microservices from Aug. 15 2022. A total of 840 upgrades were recorded. These three methods exhibit satisfactory detection performance, with precision and recall both surpassing $90\%$. Additionally, when deployed on V100 GPUs, Humas utilizing $LSDD$ exhibits significantly reduced average detection overhead, achieving a reduction of about $13.8\times$ and $2,157\times$ compared with $MMD$ and $CMMD$, respectively. This reduction is attributed to the computational simplicity involved in estimating $LSDD$ \cite{LSDDINV}. Therefore, considering the balance between precision and overhead, LSDD is a more suitable choice for online drift detection in large-scale data centers. 

It is noteworthy that 57 upgrades remain undetected by Humas, as Humas tends to disregard upgrades that do not result in noticeable drifts in CPU usage patterns.  Figure \ref{fig:overall.e} shows the CDF of the CPU usage change rate caused by the detected upgrades and the undetected upgrades, respectively. The analysis reveals that, for 98\% of the undetected upgrades, the absolute values of their CPU usage change ratios are less than 5\%. These upgrades do not lead to pattern drifts, thereby exerting minimal impact on capacity estimation.

\begin{table}[tb]
\centering
\caption{ Pattern drift detection performance of Humas with diverse parameter settings }
\label{tab:drift_perform_tune}
\vspace{-2mm}
\resizebox{0.35\textwidth}{!}{
\begin{tabular}{|c|c|c|c|c|c|}
 \hline
   \multirow{2}*{Group}&\multicolumn{3}{c|}{Parameters}&\multicolumn{2}{c|}{Overall Performance}
  \\
  \cline{2-6}
    &\multicolumn{1}{c|}{$W$}&\multicolumn{1}{c|}{$S$}&\multicolumn{1}{c|}{$\theta$}&\multicolumn{1}{c|}{Precision (\%)}&\multicolumn{1}{c|}{Recall (\%)}\\
    \cline{2-6}
  \hline
   \multirow{1}*
   {Default}&\multirow{1}*{$\mathbf{48}$}&{$\mathbf{8}$}&{$\mathbf{3}$}&{$94.40$}&{$91.04$}\\
   \cline{1-6}
   \cline{1-6}
   \multirow{2}*{Tune $W$}&\multirow{1}*{${24}$}&{${8}$}&{${3}$}&$93.21$&$79.98$\\
   \cline{2-6}
   &\multirow{1}*{${72}$}&{${8}$}&{${3}$}&$92.61$&$91.42$\\
    \cline{1-6}
    \cline{1-6}
     \multirow{2}*{Tune $S$}&\multirow{1}*{${48}$}&{${4}$}&{${3}$}&$94.04$&$90.18$\\
     \cline{2-6}
   &\multirow{1}*{${48}$}&{${24}$}&{${3}$}&$92.26$&$78.28$\\
   \cline{1-6}
    \cline{1-6}
  \multirow{2}*{Tune $\theta$}&\multirow{1}*{${48}$}&{${8}$}&{${1}$}&$96.90$&$72.36$\\
     \cline{2-6}
   &\multirow{1}*{${48}$}&{${8}$}&{${2}$}&$94.88$&$76.34$\\
   \hline
\end{tabular}
}
\vspace{-2mm}
\end{table}

\paragraph{Detection Parameters}
\label{sec:parameter_tuning}
Table \ref{tab:drift_perform_tune} presents the detection performance of Humas after performing parameter tuning on $W$, $S$ and $\theta$. The results show that our default setting outperforms other control groups in achieving a favorable trade-off between detection precision and the reduction of false positives (i.e., improving the recall ratio). This is because the default settings are based on the behaviors of microservices. Specifically, we set $W=48$ hours, considering that workload patterns of microservices generally exhibit daily periodicity \cite{kae-informer}, and 95\% of upgrade intervals exceed 2 days. Moreover, as 95\% of upgrade durations fall below 8 hours, we set $S=8$ hours. Furthermore, setting $\theta=3$ ensures Humas tolerates temporary interferences, mitigating the issue of false detection.

\paragraph{Detection Mode}
Based on accurate drift detection, Humas enhances the learning capability of the latest CPU usage patterns compared with methods employing fixed intervals. As shown in Table \ref{tab:abs_overall}, compared with $\text{Hmas}_{D8}$ and $\text{Hmas}_{D2}$, Humas improves overall resource efficiency by about 12.8\% and 14.7\% and enhances performance stability by about 18.0\% and 9.1\%, respectively. This improvement can be attributed to two factors. Firstly, unlike $\text{Hmas}_{D8}$ which adopts a longer update interval, Humas effectively eliminates outdated resource usage patterns.  
Secondly, temporary drifts caused by resource interference commonly arise in production. The excessively small update interval in $\text{Hmas}_{D2}$ may capture temporary patterns and consequently diminish the accuracy of long-term estimation. In contrast, Humas efficiently tolerates temporary pattern changes, thereby avoiding the unnecessary relearning of patterns. 


\subsection{Overhead Analysis}
\label{sec:gen_eval}



\begin{table}[tb]
\centering
\caption{ {Average overhead of modules in pattern analysis}}
\label{tab:overhead}
\vspace{-2mm}
{
\resizebox{0.35\textwidth}{!}{
\begin{tabular}{|c|c|}
 \hline
   {{Module}}&{{Avg. Overhead (s)}}
  \\
  \hline
   \multirow{6} *{} 
   {Resource normalization} &{$0.0836$}\\
    \cline{1-2}
{LSDD-based drift detection}&{$0.0146$}\\
    \cline{1-2}
    {GRF-based pattern learning}&{$2.316$}\\
    \cline{1-2}
   \hline
\end{tabular}
}
\vspace{-4mm}
}
\end{table}

{
In this subsection, we analyze the impact of the overhead incurred by Humas.
Table \ref{tab:overhead} presents the average time (s) of each module in one procedure of performance pattern learning for a microservice. It can be observed that both the resource normalization and the LSDD-based detection overheads are at the millisecond level. When compared to the hour-level sliding-window step and adjustment interval (i.e, $S=8$ hour and $h_p=1$ hour by default), we consider the overheads of pattern analysis in Humas have a negligible impact on the throughput and performance of the production microservices.} 

{
Although training GRF models to capture new performance patterns incurs a few seconds of overhead, we consider this to be acceptable due to the typical day-level upgrade interval of microservices. Furthermore, the re-training of the models is only triggered by the identification of new upgrades. Additionally, the pattern learning process for different microservices can be effectively parallelized in our implementation, thereby further enhancing the deployability of Humas.}

\section{Related Work} 
\label{related-model}


Based on recently published surveys \cite{survey1,survey2}, auto-scaling frameworks for microservices can be broadly categorized into reactive and proactive approaches.

\subsection{Reactive Approaches} 
Capacity adjustments in reactive approaches are typically triggered by performance anomalies. For example, a plethora of reactive approaches \cite{autoscaleopt,amazon} rely on rules predefined on different metrics, e.g., CPU utilization, tail latency, and QoS violation, to adjust resources. In addition, some studies \cite{FIRM,PARTIES} employ machine learning to identify the underlying cause of anomalies and generate appropriate adjustment plans. For instance, FIRM \cite{FIRM} uses reinforcement learning (RL) models to detect microservices that experience capacity under-provisioning when end-to-end QoS violations occur, subsequently scaling their resource to alleviate these anomalies. However, these methods are limited by their inability to anticipate future resource requirements, thereby failing to prevent potential service anomalies and compromising performance stability.

\subsection{Proactive Approaches} 
Numerous proactive auto-scaling approaches have been proposed, aiming to maintain performance stability by capturing performance patterns and adjusting resource capacity based on the learned patterns and the predicted workload. On one hand, several studies \cite{sinan,GRAF,ASARSA} focus on QoS satisfaction. For example, GRAF \cite{GRAF} and Sinan \cite{sinan} utilize graph neural networks (GNN) and convolution neural networks (CNN) respectively to capture latency patterns and prevent tail latency overuns. On the other hand, many studies \cite{deep-scaling, Meta,facebook} concentrate on learning CPU usage patterns to enhance CPU utilization stability. For instance, DeepScaling \cite{deep-scaling} and Meta \cite{Meta} design a deep probabilistic network and a DNN model respectively to represent CPU usage patterns. 


To mitigate the impact of performance pattern drifts, on one hand, a majority of existing research \cite{deep-scaling,Meta,rzadca2020autopilot} employ an empirical approach to re-learn the performance patterns within a predefined time interval. For example, DeepScaling \cite{deep-scaling} carries out model updating every two weeks. On the other hand, in some prior studies like Sinan \cite{sinan} and Cushion \cite{zhou2022cushion}, the update of pattern learning is triggered by manual upgrade notifications from developers. However, there is a notable absence of effective mechanisms to characterize and identify the pattern drifts induced by frequent upgrades in auto-scaling approaches. This deficiency significantly compromises the accuracy of capacity estimation in production settings.


To address the performance disparity among heterogeneous machines, numerous resource normalization schemes have been proposed in prior work \cite{sriraman2019softsku,rzadca2020autopilot,borg,WSC}. For example, Google \cite{borg} normalizes the CPU cores of different machine types to a specific quantify of Google compute units (GCU), which offer equivalent computational power. However, the variation in computational power among machines is determined offline using benchmark applications and remains static for diverse microservices that execute various business logic. In addition, some studies \cite{sriraman2019softsku,WSC} realized dynamic resource normalization. For instance, Dev et al. \cite{WSC} devised a dynamic GCU mechanism that measures the fluctuating performance disparity across diverse microservices. Nevertheless, these studies rely on micro-architectural performance metrics, e.g. CPI \cite{zhang2013cpi2}. Given that CPI is reliant on workload intensity, these measurement methods may be inappropriate unless the reference workload intensity is specified \cite{yi2020cpi}.

As presented in Table \ref{tab:0}, Humas distinguishes itself from previous research in three aspects. Firstly, Humas introduces a method to effectively measure the disparity in work efficiency among diverse microservices operating on heterogeneous hardware under varying workloads. Secondly, Humas devises an online mechanism to accurately identify pattern drifts caused by version upgrades, ensuring the accurate learning of resource patterns. Thirdly, Humas is characterized by its lightweight nature, enabling efficient adaption to numerous microservices that undergo frequent upgrades in data centers.

\begin{table}[btp]
    \centering
    \caption{Comparison of existing work}
  \label{tab:0}
  \resizebox{0.475\textwidth}{!}{
    \begin{threeparttable}
    \begin{tabular}{|c|c|c|c|c|c|c|c|}
   \hline
   {\multirow{3} *{Type}}&\multirow{3} *{Approach}&\multirow{3} *{ Metric}& \multicolumn{3}{c|} {Heterogeneity Awareness}&\multicolumn{2}{c|}{Pattern Representation}\\
   \cline{4-8}
   {}&{}&{}&{Difference}&{\multirow{2}*{Online}}&{\multirow{2}*{Dynamic}}&{\multirow{2}*{Model}}&{\multirow{2}*{Upgrade-aware}}\\
   {}&{}&{}&{Measurement}&{}&{}&{}&\\
   \cline{1-8}
         \multirow{3}*{Reactive}&AutoScaleOpt \cite{autoscaleopt}&{CPU Util.}&\XSolidBrush&\XSolidBrush&\XSolidBrush&\XSolidBrush&\XSolidBrush\\
         \cline{2-8}
          &GCE\cite{googleauto}&CPU Util.&$\checkmark$&\XSolidBrush&\XSolidBrush&\XSolidBrush&\XSolidBrush\\
         \cline{2-8}
         &FIRM\cite{FIRM}&Tail Latency&\XSolidBrush&\XSolidBrush&\XSolidBrush&RL&\XSolidBrush \\ 
         \cline{1-8}
         \multirow{6}*{Proactive}&{Sinan}\cite{sinan}&Tail Latency&$\checkmark$&\XSolidBrush&\XSolidBrush&CNN &$\checkmark$,$UN$\\
         \cline{2-8}
    &GRAF&Tail Latency&\XSolidBrush&\XSolidBrush&\XSolidBrush&GNN & \XSolidBrush\\
    \cline{2-8}
    &A-SARSA\cite{ASARSA}&Tail Latency&\XSolidBrush&\XSolidBrush&\XSolidBrush&DNN & \XSolidBrush\\
    \cline{2-8}
    & Autopilot \cite{rzadca2020autopilot}&CPU Util.&$\checkmark$&\XSolidBrush&\XSolidBrush& Histogram&\XSolidBrush\\
    \cline{2-8}
    & DeepScaling \cite{deep-scaling}&CPU Util.&\XSolidBrush&\XSolidBrush&\XSolidBrush& DNN &\XSolidBrush\\
    \cline{2-8}
    & Meta \cite{Meta}&CPU Util.&\XSolidBrush&\XSolidBrush&\XSolidBrush& ANN &\XSolidBrush\\
    \cline{2-8}
    & \textbf{Humas} (ours)&CPU Util.&$\checkmark$&$\checkmark$&$\checkmark$& GRF &$\checkmark$,$DD$\\     
    \cline{1-8} 
    \end{tabular}
    \begin{tablenotes}
  \item [1] The Metric column indicates the performance metric, where CPU Util. stands for CPU utilization.
   \item [2] $DD$ and $UN$ represents the drift detection and upgrade notification, respectively.
\end{tablenotes}
    \end{threeparttable}
}
\vspace{-5mm}
\end{table}




\section{Discussion}
\label{sec:discuss}
{While Humas is capable of achieving precise resource estimation, it exhibits three limitations that can be further optimized. Firstly, the reactive drift detection approach deployed by Humas may compromise the timeliness of performance pattern learning. To address this, our future work will concentrate on investigating either predictive techniques or code update detection methods \cite{code_update_detect} to proactively anticipate pattern drifts. By doing so, we aim to facilitate early updates of performance patterns, thereby enhancing system stability.}  

{
Secondly, the precision of workload prediction plays a crucial role in proactive capacity adjustment. Previous research \cite{kae-informer} has demonstrated that the RPS time-series of different microservices exhibit varying levels of predictability in terms of growth trends, periodicity, temporal dependencies, and other factors. Therefore, in our future work, we plan to employ long-term forecasting models \cite{kae-informer,zhou2021informer} for sequences that exhibit long-range dependencies. Conversely, for microservices with highly fluctuating RPS, we will utilize short-term predictors \footnote{E.g., Crane with DSP algorithms, which can be accessed at: https://github.com/gocrane/crane} and shorten the adjustment interval to ensure precise capacity estimation.}

{
Thirdly, the overhead associated with re-learning the performance patterns of microservices, as discussed in Section \ref{sec:gen_eval}, has the potential to impede the scalability of Humas when facing frequent service upgrades. Our future work will focus on employing more lightweight models or accelerating the model training process. 
} 

\section{Conclusion}
Accurate  adjustment of resource capacity for diverse microservices poses challenges in complex production data centers. In light of this, we analyzed the two factors hinder the performance learning for microservices namely 1) the upgrade behaviors and the pattern drifts incurred by them and 2) the efficiency difference among heterogeneous machines. Motivated by the insights, we propose Humas, a solution that explicitly addresses the impact of hardware heterogeneity and version upgrades on the auto-scaling of microservices.
Through extensive experiments conducted on a set of $50$ large-scale microservices, we demonstrate that Humas achieves state-of-the-art performance stability and resource efficiency. 


\bibliographystyle{IEEEtranS}
\bibliography{res}

\vspace{11pt}


\vfill

\end{document}